\documentclass[12pt]{iopart}
\usepackage{iopams}
\usepackage{graphicx}
\newcommand{\be}{\begin{equation}}
\newcommand{\ee}{\end{equation}}
\newcommand{\sump}{\mathop{{\sum}'}_{n=0}^\infty}
\eqnobysec
\begin{document}
\title[Vacuum energy in conical space]{Vacuum energy in conical space with additional boundary conditions}
\author{V V  Nesterenko$^1$ and I G Pirozhenko$^{1,2}$}
\address{$^1$ Bogoliubov Laboratory of Theoretical Physics,
Joint Institute for Nuclear Research,
Dubna 141 980, Russia}
\address{$^2$ Dubna University, Dubna 141 980, Russia}
\eads{\mailto{nestr@theor.jinr.ru}, \mailto{pirozhen@theor.jinr.ru}}
\begin{abstract}
Total vacuum energy of some quantized fields in conical space with
additional boundary conditions is calculated. These conditions are
imposed on a cylindrical surface which is coaxial  with the symmetry
axis of conical space. The explicit form of the matching conditions
depends on the field under consideration. In the case of
electromagnetic field, the perfectly conducting boundary conditions
or isorefractive matching conditions are imposed on the cylindrical
surface. For a massless scalar field, the semi-transparent
conditions ($\delta$-potential) on the cylindrical shell  are
investigated. As a result,  the total Casimir energy of
electromagnetic field and scalar field, per a unit length along the
symmetry axis, proves to be finite unlike the case of an infinitely
thin cosmic string. In these studies the spectral zeta functions are
widely used. It is shown briefly how to apply this technique for
obtaining the asymptotics of the relevant thermodynamical functions
in the high temperature limit.
\end{abstract}
\date{\today}

\maketitle
\section{Introduction}

Quantum matter fields on manifolds with conic singularity are of
interest in black hole physics and especially in cosmic string
theory. The status of cosmic strings was changing during their
rather long history. Therefor it is worth mentioning a few words
about this topic.

 In the course of time evolution and extension, the
Universe should undergo  a number of phase transitions. From quite
general point of view, these transitions may be accompanied by
emergence of topological defects\footnote{The topological defects
are the  regions with higher dynamical symmetry surrounded by a
space with lower symmetry group.} such as  monopoles, cosmic
strings, and domain walls \cite{Kibble-1}. When neglecting the
internal structure of these objects, we are left with point-like,
one- and two-dimensional defects, respectively. For the cosmic
scenarios the most interesting were monopoles and cosmic strings. On
this subject, there is a vast literature (see, for instance,
\cite{Vilenkin-1} and references therein). In the 1980's, the cosmic
strings were considered as candidates for a mechanism of galaxy
formation. However, recent observations on the cosmic microwave
background have shown conclusively that this effect can at most
account for a small fraction of the total power, up to 10
\%~\cite{CMB-1,CMB-2}. Nevertheless, the cosmic strings are still
considered as the plausible sources of detectable gravitation waves
(see, for example, \cite{GW-1,GW-2} and references therein), gamma
ray burst \cite{GRB}, high-energy cosmic rays \cite{BCXZ}, and due
to their gravitational lensing effects \cite{lensing}. Recently,
cosmic strings attract a renewed interest in the framework of brane
inflation \cite{brane} and as the macroscopic fundamental strings
\cite{Kibble-2,Kibble-3,Pol,cosmic_ss}. Comprehensive review of the
cosmic string studies are presented in papers
\cite{Vilenkin-2,Kibble-4}.

The gravitational field, i.e.\ geometry of the space-time, connected
with cosmic string, is determined by the solution of the Einstein
equations with the energy-momentum tensor defined by the string.
When infinitely thin straight string with the linear mass density
$\mu $ is located along the $z$ axis then the space outside the
string is isomorphic to the manifold $ {\mathbb R}^1 \times C_{2\pi
- \phi}$, where ${\mathbb R}^1$ is an infinite line along the $z$
axis, and instead of a plane perpendicular to this axis, one has two
dimensional cone $C_{2\pi-\phi}$ with the angle deficiency $\phi =8
\pi \mu G $, $G$ being the gravitational constant. The cone
$C_\beta$  of an angle $\beta $ becomes a plane when $\beta =2\pi$.
The static straight thin cosmic string is thus represented by a
locally flat space-time with a conical singularity along an axis
\cite{Vilenkin_cone,Hiscock,Linet,Starobinsky}. A string with a
deficit angle has a positive mass density and it is stretched; a
string with an excess angle has negative mass density and is
squeezed. Beside the empty space-time with a single string, cosmic
strings appear in a wide variety of solutions to Einstein equations.
Any axially symmetric space-time can be trivially modified to
contain a string on the axis of symmetry. Keeping in mind the
physical origin of a cosmic string, one should characterize it by
the gauge flux parameter in addition to the deficit angle parameter.
However this point is beyond the scope of our consideration.

For physical applications  it is interesting to study
different fields in the background of cosmic string. As
explained above, this problem is reduced  to
consideration of the fields on the space-time
manifold\footnote{The manifold ${\mathbb R} ^2\times
C_\beta$ arises under consideration of fields on the
Rindler space-time at finite temperature $\beta ^{-1}$.}
${\mathbb T}^1\times {\mathbb R}^1 \times C_{2\pi - \phi}$.
It is assumed that the effect of the fields  on the
space-time geometry can be neglected.

One of the difficulties  in these studies is nonintegrable
singularity at the string axes for the energy density of fields
considered in the background of infinitely thin cosmic string
\cite{NLS-AP}. As a result, the total vacuum energy (per unit length
of the string),  is not defined. To overcome this drawback,  the
cosmic string of finite thickness can be considered. For this
purpose one may use, for example, the Gott-Hiscock metric
\cite{Hiscock,Gott} which is the solution to the Einstein equations
when the energy density is a constant inside the string of a finite
thickness and zero outside of it~\cite{KB}.

In the present paper, we are going to draw the attention to the
following interesting fact that is closely related to the problem
stated above. Namely, in order to render the total field energy in
conical space to be finite it is sufficient to impose some boundary
(or matching) conditions for quantized fields on a cylindrical
surface which is coaxial with the symmetry axes of the conical
space. In a loose form, this observation can be interpreted as
follows: isolation of a `naked' conic singularity by introducing
additional boundary conditions on a surface surrounding it results
in `improving' the physical consequences generated by this
singularity.

When choosing the fields for calculating the vacuum energy, we take
into consideration the following. The vacuum energy of massless
fields is, as a rule, much greater than that of massive fields.
Therefore we study the Casimir energy of electromagnetic and
massless scalar field in the problem in question. As regards the
boundary or matching conditions on the cylindrical surface,  there
is a large freedom here. We impose such conditions which  are
usually considered in the analogous studies \cite{book,Milton}. In
the case of electromagnetic field, the perfectly conducting boundary
conditions or isorefractive matching conditions are widely used in
the Casimir calculations. In our consideration, we follow in this
way. For massless scalar field, we impose the semi-transparent
conditions on the cylindrical shell. The point is that real
boundaries cannot confine the physical fields exactly within a given
region, i.e.\ real boundaries are  always more or less transparent
for fields. This property  is simulated, for example, by the
$\delta$-potential~\cite{Milton}.

As a result,  the total Casimir energy of the  electromagnetic field
and the scalar field, per a unit length along the symmetry axis, in
the conical space with additional boundary conditions proves to be
finite unlike the case  of an infinitely thin cosmic string. In our
studies of vacuum energy, we use substantially the spectral zeta
function technique. In addition, this approach enables us to
calculate directly the asymptotics of the relevant thermodynamical
functions in the high temperature limit.

The layout of the paper is as follows. In section 2, the spectra of
an electromagnetic field and a scalar massless field in the geometry
under consideration are analysed. For electromagnetic field, two
types of the boundary conditions are considered: perfectly
conducting conditions and isorefractive ones. For a massless scalar
field, semi-transparent matching conditions are applied, i.e.\ the
$\delta$-potential at the point $r=a$ is introduced,  both signs of
the respective coupling constant being investigated in detail. In
section 3, the global zeta functions for these spectra are
constructed by implementing the analytical continuation needed.
Proceeding from these results, the finite and unique values for the
vacuum energy of the electromagnetic field and scalar massless field
in the conical space with additional boundary conditions are
derived. In section 4, the high temperature asymptotics of the
relevant thermodynamical functions in the problem under study are
obtained by making use of the heat kernel coefficients calculated in
our previous work \cite{NPD} and in the present paper. In section 5,
we discuss briefly the obtained results and compare them with
analogous studies.

\section{The excitation spectrum of the fields  in conical a space with additional
boundary conditions}
When the energy-momentum tensor is determined by the $\delta$-like
mass distribution along the $z$-axis with the linear density $\mu$
the solution to the Einstein equations is given by the following
metric \cite{Vilenkin_cone,Hiscock,Linet,Starobinsky}:
\begin{equation}
\label{2-1} \rmd s^2=-\rmd t^2+\rmd r^2+\left (1-4\mu G \right )^2
r^2\rmd \varphi^2 +\rmd z^2, \quad 0\leq \varphi \leq 2\pi\,{,}
\end{equation}
where $G$ is the gravitational constant. The cylindrical coordinates
$(r,\varphi,z)$ are used with the $z$-axis coinciding with thin
straight infinite string. It is easily convinced that the space part
of  metric (\ref{2-1}) describes the geometry on the surface of a
cone with the deficit angle $\Phi=8\pi  \mu G $ or with the opening
angle $2\pi -\Phi$.

When considering matter fields, on the cone surface one can
obviously substitute the latter by a plane with a segment of the
angle $\Phi $ cut out, provided the periodicity conditions in a new
angular variable $\theta$
\begin{equation}
 \label{2-2}
f(\theta)=f(\theta +2 \pi - \Phi), \quad 0 \leq \theta \leq2\pi -\Phi
\end{equation}
 are imposed.

In terms of the angular variable $\theta$  metric (\ref{2-1}) looks
exactly like the metric of the Minkowski spacetime
\begin{equation}
\label{2-1a} \rmd s^2=-\rmd t^2+\rmd r^2+
r^2\rmd \theta^2 +\rmd z^2
\end{equation}
supplemented by  conditions (\ref{2-2}) (see for details
\cite{NPD}).

\subsection{Electromagnetic field}
In the case of an  electromagnetic field in a conical   space, we
apply perfectly conducting conditions or isorefractive matching
conditions on an auxiliary cylindrical surface. As a result, the
problem under consideration is reduced to the calculation of the
electromagnetic vacuum energy for a cylindrical shell by imposing on
its lateral surface boundary conditions (or matching conditions)
mentioned above and  by substituting the usual $2\pi$-periodicity by
the periodicity condition (\ref{2-2}).

The solutions to the Maxwell equations with boundary conditions on
the circular cylindrical shell have been considered in many papers
(see, for example, \cite{NLS-AP,NPD,NLS}). We shall take advantage
of this analysis and allow for  the periodicity condition
(\ref{2-2}). The latter requirement leads to the substitution of the
integer index $n$   in the Bessel and Hankel functions  by $n p$,
where \be \label{2-3} \frac{1}{p}=1 - \frac{\Phi}{2 \pi}\,{.} \ee
According to the physics of  cosmic string formation
\cite{Vilenkin-1,Kibble-2,Vilenkin-2}, $p$ is very close to one and
$p>1$.

Let us address now  the spectrum of electromagnetic oscillations in
the configuration under consideration. In view of the  axial
symmetry of the conical space the eigenvalues of the spectral
problem (eigenfrequencies) $\omega_q $ are `numbered' by the
following three indexes:
 \be \label{2-4}
q\equiv \{n_r,n,k_z \}, \ee which correspond to the coordinates
$\{r,\theta,z\}$. The radial index $n_r$ can take discrete values or
it can be continuous. The angle index $n$ corresponds to the compact
angle variable $\theta$: $0\leq \theta \leq 2\pi -\Phi= 2\pi/p$. The
continuous variable $k_z$ in (\ref{2-4}) is the wave vector of the
plane waves propagating in $z$-direction: $-\infty <k_z< \infty $.

General solution to Maxwell equations in  metric (\ref{2-1a}) in
unbounded space-time is expressed in terms of two independent scalar
functions that correspond to the TE-polarization and TM-polarization
of the electromagnetic field \cite{NVV-sm,Whittaker,Nisbet}. In the
conical space, these functions should satisfy the periodicity
conditions (\ref{2-2}).

For further specification of the spectrum in the problem at hand,
the boundary conditions on the cylindrical shell should be taken
into account explicitly.

\subsection{\label{2a}Perfectly conducting boundary conditions}
On the surface of a perfect conductor, the tangential component of
electric field $(\mathbf{E}_{\parallel})$ and the normal component
of the magnetic field $(\mathbf{H}_{\perp})$ should vanish
\begin{equation}
\label{2-4a}
\mathbf{E}_{\parallel}=0,\quad \mathbf{H}_{\perp}=0\,{.}
\end{equation}
It is important to note that these conditions do not couple
the polarizations of electromagnetic fields, as the Maxwell
equations do,  and furthermore the
fields inside and outside of the shell are independent~\cite{Stratton}.

Inside the shell, the (unnormalized) eigenfunctions are~\cite{NPD}
\begin{equation}
\label{2-5}
u_{nn_r}(r,\theta)= J_{np}(\lambda_{nn_r}r)
{{\sin np\theta}\choose{\cos np\theta}}, \quad n=0,\,1,\,2,\,\ldots\,{,}
 \end{equation}
where the parameter $p$ was  introduced in  (\ref{2-3}) and
\be
\label{2-6}
\lambda^2=\frac{\omega^2}{c^2}-k^2_z\,{.}
\ee
For simplicity we drop here the common multiplier
$\exp{(-\rmi\omega t+\rmi k_z z)}$.
Solutions to the Maxwell equations are expressed in terms of the eigenfunctions (\ref{2-5})
through the well known formulas \cite{NLS,NVV-sm}.

For the TE-polarization, $\lambda_{nn_r}$ in (\ref{2-5}) is  the
$n_r$-th root  of the equations
\begin{equation}
\label{2-8}
J'_{np}(\lambda_{nn_r}a)=0,\quad n=0,\,1,\,2,\,\ldots, \quad n_r=1,2, \ldots\,{,}
\end{equation}
and for the TM-polarization, we should take the roots of the
equations
\begin{equation}
\label{2-9}
J_{np}(\lambda_{nn_r}a)=0,\quad n=0,\,1,\,2,\, \ldots, \quad n_r=1,2, \ldots \,{.}
\end{equation}

The explicit form of the eigenfunctions (\ref{2-5})
implies, specifically, that at $n\neq 0$  all the
eigenvalues in the spectral problem under consideration are
twice degenerate (the both functions $\sin np\theta$ and
$\cos np\theta$ work) and at $n=0$ this degeneracy is
eliminated because $\sin np\theta$ vanishes.

Summarizing we can infer the following: At fixed value of $k_z$ the
spectrum of the frequencies $\omega(\mathrm{TE},k_z)$ and
$\omega(\mathrm{TM},k_z)$ inside the shell is discrete and it is
given by the formula \be \label{2-10}
\omega=c\sqrt{\lambda^2+k_z^2}\,{,} \ee where
\[
\omega=\omega_{nn_r}(\sigma,k_z),\quad \sigma=\mathrm{TE},\;\mathrm{TM}\,{,}
\]
and for the TE-modes, $\lambda$ in (\ref{2-10}) should be
substituted be the roots of the equation (\ref{2-8}) and in the case
of  the TM-modes the roots of the   equation (\ref{2-9}) should be
used.

Outside of the shell the spectrum of the electromagnetic
oscillations is continuous
\begin{equation} \label{2-10a} ck_z<\omega
<\infty\,{,} \end{equation} because we are dealing here
with free unbounded configuration space. For the summation
over such a spectrum to be accomplished the scattering
formalism should be employed \cite{NVV-bc}, i.e.\ one has
to calculate the pertinent $S$-matrix. The regular
solutions defining the Jost functions $a^\pm_n(\lambda)$ in
our model read \begin{equation}
 \label{2-11}
f_{n}(r)=a_n^-(\lambda)H^-_{np}(\lambda r)-
a_n^+(\lambda)H^+_{np}(\lambda r),\quad r>a\, {,}
\end{equation}
 where
$\lambda$ is a continuous variable related to $\omega $ and
$k_z$ by  (\ref{2-6}) and $H_n^\pm(\lambda r)$ are the
Hankel functions
$$
H_n^+(\lambda r)\equiv H_n^{(1)}(\lambda r), \quad H_n^-(\lambda r)\equiv H_n^{(2)}(\lambda r)\,{.}
$$
The regular solutions (\ref{2-11}) should satisfy the boundary conditions
\be
\label{2-12}
\left .\frac{d}{dr}f^{\mathrm{TE}}_{n}(r)\right |_{r=a}=0,\quad
\left .f^{\mathrm{TM}}_{n}(r)\right |_{r=a}=0\,{.}
\ee

By making use of the definition of the $S$-matrix
$$
S_n(\omega)=\frac{a_n^+(\lambda)}{a_n^-(\lambda)}
$$
and  (\ref{2-11}), (\ref{2-12}) we obtain
\be
\label{2-13}
S^{\mathrm{TE}}_n(\omega)=\frac{H_{np}^{-\prime}(\lambda a)}{H_{np}^{+\prime}(\lambda a)}, \quad
S^{\mathrm{TM}}_n(\omega)=\frac{H_{np}^{-}(\lambda a)}{H_{np}^{+}(\lambda a)}\,{.}
\ee

The note concerning the different degeneracy of the states
with $n=0$ and $n\neq 0$ for oscillations inside the shell
applies also to the scattering states outside of the shell.

When $\omega^2 < c^2k_z^2$ there are solutions to the wave
equation which decay in the radial direction
$$
H_{np}^+(\lambda r)\equiv H_{np}^{(1)}\left(i\frac{r}{c}\sqrt {c^2 k_z^2-\omega^2}
\right )=\frac{2}{\pi}i^{-n-1}K_{np}\left(\frac{r}{c}\sqrt {c^2k_z^2-
\omega^2}
\right ){.}
$$
However they do not meet the boundary conditions
(\ref{2-12}). Hence in the problem at hand there are no
surface (or evanescent) waves.

\subsection{Isorefractive (Diaphanous) matching conditions}
\label{2b}
When employing these conditions we assume that the electric and magnetic
properties of the media inside ($\varepsilon_1,\mu_1$) the
shell and outside of it ($\varepsilon_2,\mu_2$) are
different, but the velocities of light  in these regions
are the same: $c_1=c_2,\;
c_i=c/\sqrt{\varepsilon_i\mu_i},\;i=1,2$ ($c$ is the light
velocity in vacuum). The fields $\mathbf{E}$ and
$\mathbf{H}$ inside the shell and outside of it are coupled
due to the continuity requirement for their tangential
components
\be \label{2-14} \mathrm{discont}\;
(\mathbf{E}_{\parallel}) =0,\qquad \mathrm{discont}\;
(\mathbf{H}_{\parallel})=0, \qquad r=a\,{.}
\ee

 The configuration space is noncompact, and as a
consequence the electromagnetic spectrum is continuous
(\ref{2-10a}). Indeed, for a given $n$ and $k_z$ the scattering
solutions to the Maxwell equations contain 6 amplitude: the
TE-solution has 1 internal amplitude and 2 amplitudes outside of the
shell. The same holds for the TM-solution. The matching conditions
(\ref{2-14}) at the cylinder surface lead to  4 linear homogeneous
equations for these amplitudes. Hence no restrictions arise here for
the spectral parameter $\omega^2/c^2$. The polarizations in this
problem decouple in the equations determining the eigenfrequencies
\cite{MNN}, but they do not decouple on the level of the scattering
solutions.\footnote{This is also true for the natural modes of a
compact cylinder with $c_1=c_2$  (see (6) and (7) in section 9.15 of
the book \cite{Stratton}). This set of equations does not split into
equations separately for $a_n^e,\;a_n^e$ and for  $b_n^e,\;b_n^e$,
when $c_1=c_2$.} However, the scattering matrix in this case has a
simple structure (more precisely, the Jost matrices have zero
diagonal elements). In our paper \cite{NVV-bc} the $S$-matrix has
been derived for the scattering of electromagnetic waves on an
infinite circular material cylinder in the general case when
$c_1\neq c_2$. These formulae are considerably simplified if
$c_1=c_2$. As before, the $S$-matrix is the $(2\times 2)$ matrix
acting on  polarizations (TE and TM) and it satisfies the  matrix
equation\footnote{It is convenient to interchange the notations for
the matrices $K^{\pm}$ in \cite{NVV-bc}: $K^{\pm}\to K^{\mp}$ and
take in the right-hand side of  (20) in \cite{NVV-bc} the difference
instead the sum in order to comply with  (\ref{2-11}) in the present
paper.} \be \label{2-16} K^+ S+K^-=0 \, {,}\ee where the matrices
$K^{\pm}$ are  now \be
 \label{2-17}
K^{\pm}=\pm\left (
\begin{array}{cc}
0 & \beta^\pm \\
\gamma^\pm& 0
\end{array}
\right ){,}
\ee
\[
\beta^{\pm}_{np}=-\rmi\frac{\omega \lambda}{cJ_{np}}\left (
\mu_2J_{np}{H^{\pm}_{np}}' -\mu_1 {J_{np}}'{H^{\pm}_{np}}
\right ),
\]
\[
\gamma^{\pm}_{np}=\rmi\frac{\omega \lambda}{cJ_{np}}\left (
\varepsilon_2J_{np}{H^{\pm}_{np}}' -\varepsilon_1 {J_{np}}'{H^{\pm}_{np}}
\right ){.}
\]
In these formulae, all the Bessel ($J_{np}$) and Hankel
($H^{\pm}_{np}$) functions have the same argument $a\lambda$, where
$\lambda = \sqrt{(\omega/c)^2-k_z^2}$. Ultimately, the determinant
of the $S$-matrix
 \begin{equation} \label{2-18} \det
S=-\frac{\det K^-}{\det K^+} \end{equation} is expressed in terms of
the known multipliers $\Delta_n^{\mathrm{TE}}$ and
$\Delta_n^{\mathrm{TM}}$ \cite{NLS,MNN,LNB}, these multipliers being
constructed for  the both Hankel functions of the first
($H_{np}^{(1)}$) and  second ($H_{np}^{(2)}$) kinds.\footnote{In
\cite{LNB} (see endnote 18 there), it was proposed to do this `by
hand' without rigorous justification. In  \cite{MNN} rather
complicated contours were used  when gong to the imaginary
frequencies.}

It is easy to show that in the problem under consideration
there are no surface modes. For $\omega^2<c^2k_z^2$ there
are solutions to the wave equation which decay in both
directions from the lateral surface of the cylindrical
shell: $u_n(r) \sim I_{np}(r\sqrt{c^2k_z^2-\omega^2}/c), \;
r<a$ and $u_n(r) \sim K_{np}(r\sqrt{c^2k_z^2-\omega^2}/c),
\; r>a$. The respective frequency equations are
\begin{equation}
\label{2-19}
\varepsilon_1 \frac{{I_{np}}'}{I_{np}} - \varepsilon_2 \frac{{K_{np}}'}{K_{np}}, \quad
\mu_1 \frac{{I_{np}}'}{I_{np}} - \mu_2 \frac{{K_{np}}'}{K_{np}} =0\,{,}
\quad n=0,1,2, \ldots\,{.}
\end{equation}
Here all the modified Bessel functions have
the same argument $(a/c)\sqrt{c^2k_z^2-\omega^2}$. Since
${I_{np}}'>0$ and ${K_{np}}'<0$ the left-hand sides of
Eqs.\ (\ref{2-19}) are strictly positive. Hence these
equations have no real roots in the interval $0<\omega <ck_z$.

Summarizing this subsection we infer that the spectrum of
electromagnetic oscillations in the case of isorefractive matching
conditions  is pure continuous (see (\ref{2-10a})).

\subsection{Semitransparent matching conditions for scalar field}
Now we address the consideration of a massless scalar field $\varphi
(t,\mathbf{x})$ in the conical space with a cylindrical surface of
radius $a$. On this surface  we impose on the field $\varphi (t,
\mathbf{x})$ the matching conditions  which may be interpreted as
the scalar $\delta$-potential. The wave equations in this case read
\begin{equation}
\label{2-20} \left [ \frac{1}{c^2}\frac{\partial ^2}{\partial
t^2}-\Delta+\frac{g}{r}\delta(r-a) \right
]\varphi(t,\mathbf{x})=0\,{,} \ee where $g$ is a dimensionless
constant specifying the strength of the potential and $\Delta$ is
the Laplace operator for the spatal part of  metric (\ref{2-1a}).
Separating the variables
\begin{equation}
\label{2-21}
\varphi(t,\mathbf{x})\sim
\rme^{-\rmi\omega t+\rmi k_z z}f_n(r){{\sin np\theta}\choose{\cos np\theta}}, \quad n=0,\,1,\,2,\,\ldots\,{,}
\ee
we arrive at the equation for the radial  function $f_n(r)$
\begin{equation}
\label{2-22} \frac{d^2f_n(r)}{dr^2}+\frac{1}{r} \frac{d
f_n(r)}{dr} + \left [ \frac{\omega ^2}{c^2}
-k_z^2-\frac{n^2p^2}{r^2}-\frac{g}{r}\,\delta(r-a)
\right]f_n(r)=0\,{.} \ee First we demand the function
$f_n(r)$ to be continuous at the point $r=a$
\begin{equation}
\label{2-23} f_n(a+0)-f_n(a-0)=0,
\ee
so that one could integrate
the left-hand side of  (\ref{2-22}) over $r \rmd r$. Such an
integration in the viicnity of the point $r=a$ yealds
\begin{equation}
\label{2-24} f'_n(a+0)-f'_n(a-0)=\frac{g}{a}f_n(a)\,{.}
\end{equation}
As usually \cite{Albeverio} the wave equation with the
$\delta $-potential (\ref{2-22}) is substituted by a pertinient free
equation (without potential)
\begin{equation}
\label{2-25}
\frac{d^2f_n(r)}{dr^2}+\frac{1}{r} \frac{d
f_n(r)}{dr} + \left ( \frac{\omega ^2}{c^2}
-k_z^2-\frac{n^2p^2}{r^2}
\right )f_n(r)=0\,{,}
\end{equation}
the solutions of the latter, $f_n(r)$, being subjected to
the matching conditions (\ref{2-23}) and (\ref{2-24}).
Sometimes the matching conditions generated by the
$\delta$-potential are refered to as the semitransparent
ones.

Finally at a given $k_z,\;-\infty<k_z<\infty$ we face  the
spectral problem specified by differential equation
(\ref{2-25}), matching conditions (\ref{2-23}),
(\ref{2-24}) and physical conditions at the origin $(r=0)$
and  at the infinity $(r\to \infty)$, the frequency
$\omega>0$ being the spectral parameter. It turns out that
the structure of the spectrum in the problem at hand
depends strongly on the sign of the constant $g$, namely,
for positive values of $g$ the spectrum is pure continuous,
and when $g<0$ we have, in addition to the continuous
branch, surface modes (or bound states).

First we consider the continuous part of the
spectrum, i.e.\  we construct the scattering matrix in
this problem for both signs of $g$.
The regular solution is
\begin{equation}
\label{2-26}
f_n(r)=\cases{a_n^{in}(\lambda) J_{np}(\lambda r)& for  $r<a$,\cr
a_n^{-}(\lambda) H_{np}^-(\lambda r)-
a_n^{+}(\lambda) H_{np}^+(\lambda r) & for  $r>a$\,{,}\cr}
\end{equation}
where
\begin{equation}
\label{2-27}
\lambda=\sqrt{(\omega/c)^2-k_z^2}\,{.}
\end{equation}
The matching conditions  (\ref{2-23}) and (\ref{2-24}) yield
\begin{equation}
\label{2-28}
a_n^{in}(\lambda)J_{np}(\lambda a)=
a_n^{-}(\lambda)H^-_{np}(\lambda a)-a_n^{+}(\lambda)H_{np}^+(\lambda a)\,{,}
\end{equation}
\begin{equation}
\label{2-29}
a^{in}_n(\lambda)\left [\frac{g}{\lambda a} J_{np}(\lambda a)+
 {J_{np}}'(\lambda a)\right ]=a_n^{-}(\lambda){H^-_{np}}'(\lambda a)-a_n^{+}(\lambda){H_{np}^+}'(\lambda a)
{.}
\end{equation}
Here and below the prime in Bessel and Hankel functions denotes the differentiation
with respect to their arguments, $\lambda a$.
Eliminating from  (\ref{2-28}) and (\ref{2-29}) the amplitudes $a_n^{in}(\lambda)$,
we obtain the $S$-matrix
\begin{equation}
\label{2-30}
S_n(\omega)=
\frac{gJ_{np}(\lambda a)H_{np}^-(\lambda a)+
{2\rmi}/{\pi}}{gJ_{np}(\lambda a)H_{np}^+(\lambda a)-{2\rmi}/{\pi}}{.}
\end{equation}

Now we address the consideration of the surface modes in
the problem under study. Such solutions to the
boundary-value problem (\ref{2-25}), (\ref{2-23}), (\ref{2-24}) exists only
for negative constant $g$. Indeed, when
\[
\frac{\omega^2}{c^2} - k_z^2<0\,{,}
\]
the radial wave equation (\ref{2-25}) has the solutions
$K_{np}(\kappa r)$ and $I_{np}(\kappa r)$, which exponentially
vanishe outside  the cylinder shell in both directions, for $r>a$
and for $r<a$ respectively. Here \be \label{2-40}
\kappa=\sqrt{k_z^2-\frac{\omega^2}{c^2}}\,{.} \ee Thus, the complete
solution describing the evanescent waves (surface modes) is given by
\begin{equation}
\label{2-41}
f_n(r)=\cases{
a_n^{in}(\kappa)I_{np}(\kappa r) & for $r<a$, \cr
a_n^{ex}(\kappa)K_{np}(\kappa r) & for $r>a${.}\cr}
\end{equation}
The matching conditions (\ref{2-23}) and (\ref{2-24}) give
\be
\label{2-42}
a_n^{in}(\kappa)I_{np}(\kappa a)=a_n^{ex}(\kappa)K_{np}(\kappa a)\,{,}
\ee
\be
\label{2-43}
a_n^{ex}(\kappa)K'_{np}(\kappa a)-a_n^{in }(\kappa)I'_{np}(\kappa a)=
\frac{g}{a\kappa}a_n^{in}(\kappa)I_{np}(\kappa a)\,{.}
\ee
Taking into account the relation \cite{AS}
\[
I_\nu(x)K'_\nu(x)-I'_\nu(x)K_\nu(x)=-\frac{1}{x}\,{,}
\]
we obtain equation determining the frequencies of the surface modes:
\begin{equation}
 \label{2-44} 1+gI_{np}(x)K_{np}(x)=0\,{,}
\end{equation}
where \be\label{2-45} x^2=a^2\kappa^2=a^2\left (
k_z^2-\frac{\omega^2}{c^2} \right){.}
\end{equation}
The product of  modified Bessel functions  $I_{\nu}(x)K_{\nu}(x)$
has the following properties
\begin{eqnarray}
 \label{2-44a}
 I_0(x)K_0(x)&\sim & \ln x, \quad x\to 0\,{,}\nonumber\\
 I_{\nu}(0)K_{\nu}(0)&= &\frac{1}{2\nu}, \quad \nu>0\,{,} \\
 I_{\nu}(x)K_{\nu}(x)& \sim & \frac{1}{2x}\left [
1-\frac{4\nu^2-1}{(2x)^2}+\cdots
\right ], \quad x\to \infty\,{,}\nonumber
\end{eqnarray}
(see, for example, \cite{AS}). By numerical
calculations\footnote{The authors are indebted to M.\ Bordag for
such calculations.} one can easily make sure that, for all $\nu$,
the product $I_{\nu}(x)K_{\nu}(x)$  monotonically decreases in the
interval $0<x<\infty$ varying between the limiting values given in
(\ref{2-44a}).
Hence, the  frequency equation (\ref{2-44}) with $g<
0$  has always a root for  $n=0$ but  for $n\geq 1$ this equation
has  the root, provided that the inequality
\begin{equation}
\label{2-46} |g|^{-1}< \frac{1}{2 n p}
\end{equation}
is satisfied. Obviously for a given $|g|$ the  condition
(\ref{2-46}) can be true only for some first values of $n$.
As a consequence, merely for such $n$ the frequency
equation (\ref{2-44}) can have solution.

Let $x=a\kappa$ be a root of (\ref{2-44}), then the  frequency of the surface mode,
$\omega_{\mathrm{sm}}$, is given by
\begin{equation}
\label{2-47}
\omega_{\mathrm{sm}} =c\sqrt{k_z^2-\kappa^2}\,{,}
\end{equation}
(see  (\ref{2-45})). Obviously, the frequency
$\omega_{\mathrm{sm}}$ should be a real quantity thus one has to
add to  (\ref{2-47}) the   restriction
\begin{equation}
\label{2-48}
k_z>\kappa\,{.}
\end{equation}

Summarizing we can claim that the spectral problem under
consideration has the  continuous spectrum in the domain
(\ref{2-10a}) for both signs of the coupling constant $g$. For
negative $g$ there appear, in addition, the surface modes with
frequencies (\ref{2-47}), (\ref{2-48}).

Closing this section it is worth mentioning  solutions
in the problem under consideration which are obtained from
the surface modes when $k_z<\kappa$. These solutions, as the surface
modes, are located around $r=a$ (see  (\ref{2-41})), but they
have pure imaginary frequencies
\begin{equation}
\label{2-49} \omega =\pm \,\rmi \,\widetilde  \omega\quad  \mathrm{with} \quad \widetilde
\omega=c\sqrt{\kappa^2-k_z^2},\quad \kappa >k_z\,{.}
\end{equation}
As a result, such solutions exponentially rise with time
$\sim\exp(\pm \,\widetilde\omega \,t)$. Hence, already at the
first-quantized level, there is instability in the system at hand.
However, in what follows we disregard this instability. In our paper
\cite{shell} analogous solution was discussed briefly in the
framework of the flat plasma sheet model with negative parameter
$q$, and it was interpreted as a resonance solution. In reality, it
is not the case and it is impossible to confront the considered
solutions of the Klein-Gordon equation with any special solution of
the Schr\"odinger equation.

\section{Global zeta functions and Casimir energies}
Now we are in position to construct the spectral zeta functions
\be
\label{3-1}
 \zeta(s)= \sum_{\{q\}}\left (\omega^{-2s}_q -\bar
\omega_q^{-2s} \right ) \ee and to calculate, on this basis,
the vacuum energy of quantum fields in question
\be
\label{3-2} E=\frac{\hbar }{2}\,\zeta \left (
s=-\frac{1}{2} \right ){.}
\ee
 The
summation in (\ref{3-1}) should be done over the whole
spectrum (discrete, continuous, and with allowance for the
surface modes if such exist). The frequencies
$\bar\omega_q$ in (\ref{3-1}) are obtained from $\omega_q$
when $a\to \infty$. The subtraction of $\bar \omega _q$ in
(\ref{3-1}) corresponds to removing from the vacuum energy
the contributions proportional to the volume of the
manifold and to the  manifold boundary. In this
way the renormalization of the vacuum energy is
accomplished. The parameter $s$ is considered at first to
belong to the region of the complex plane $s$ where the sum
in (\ref{3-1}) exists. After that the analytical
continuation of (\ref{3-1}) to the point $s=-1/2$ and
further for $\mathrm{Re}\, s<-1/2$ should be done. When
summing over the continuous branch of the spectrum, the function of
the spectral density shift have to be employed \cite{NVV-bc}.
The rigorous mathematical theory
of scattering gives the following expression for this function
\be
\label{3-3}
\Delta \rho(\omega)\equiv\rho (\omega)- \rho _0(\omega)=
\frac{1}{2\pi \rmi}\frac{\rmd}{\rmd\omega}
\det S(\omega)\,{,}
\ee
where $S(\omega)$ is the $S$-matrix in the spectral problem at hand.
Here $\rho(k)$ is the density of states for a given potential
(or for a given boundary conditions in the case of compound
media) and $\rho_0(k)$ is the spectral density in the respective
free spectral problem (for vanishing potential or for
homogeneous unbounded space).
\subsection{Perfectly conducting boundary conditions}
As  shown in section \ref{2a}, the spectrum in this
problem has two branches: discrete branch (electromagnetic
oscillations inside the  shell) and continuous branch
(oscillations outside of the shell). In a complete form  (\ref{3-1}) reads
now
\begin{equation}
\label{3-4}
\fl \zeta(s)=2\sum_{\sigma}\int\limits_{-\infty}^\infty\frac{\rmd k_z}{2\pi}
\sump\left[
\sum_{n_r} \omega _{nn_r}^{-2s}(\sigma,k_z) + \int\limits_{ck_z}^\infty
\omega^{-2s}\, \Delta \rho_n (\sigma,\omega,k_z)\,\rmd \omega \right ]-(a\to \infty){,}
\end{equation}
where $\sigma=\mathrm{TE,\,TM}$ and the prime over the sum sign means that the term with
$n=0$ is taken with the weight $\frac{1}{2}$. The discrete
frequencies $\omega _{nn_r}(\sigma,k_z)$ in (\ref{3-4}) are
defined by   (\ref{2-8}) and (\ref{2-9}), and the
function of the spectral density shift is given by
(\ref{3-3}) and (\ref{2-13}).

\noindent
\begin{figure}[th]
\noindent \centerline{
\includegraphics[width=55mm]{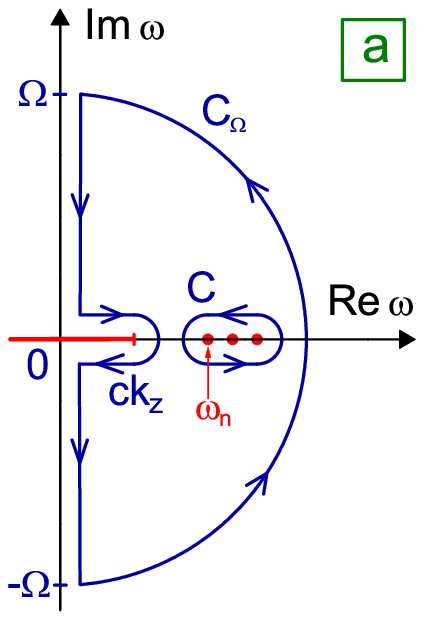}
\hspace{20mm}
\includegraphics[width=55mm]{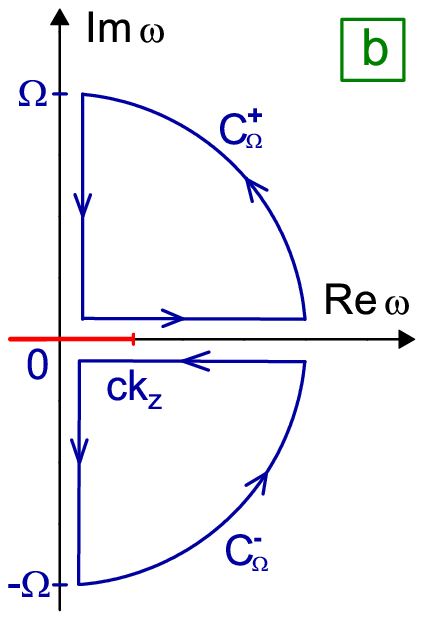}}
\caption{The contours on the complex $\omega$ plane which are used when
going on to the integration over the imaginary  frequencies in the
case of bounded (a) and unbounded (b) configuration spaces.}
\label{Plot:Contours2}
\end{figure}
The further treatment of  (\ref{3-4}) is aimed at
deriving a unique integral representation for both
contributions of discrete branch and continuous branch of
the spectrum, the integration being done over imaginary
frequencies $\omega \to \rmi\, \omega$. This is  accomplished by
transforming the sum $\sum_{n_r}$ into the contour integral
applying the argument principle theorem from the complex
analysis (see figure 1a). After that the initial contour $C$
is continuously transformed into the contour $C_\Omega$
with $\Omega \to \infty$. The contribution of the
continuous part of the spectrum is also expressed in terms
of integrals along the contours $C^\pm_\Omega$ shown in figure (1b).
The contour $C_\Omega^+\;(C_\Omega^-)$ is used for transforming the integral
$\int _{ck_z}^\infty \rmd\omega \,\omega^{-2s}\ldots$ in ({\ref{3-4}) containing
$H^+\;(H^-)$ and ${H^+}'\;({H^-}')$ in accord with (\ref{2-13}).
In both the cases the integration is accomplished on the
complex frequency plane $\omega$ with the cut connecting
the points  $ck_z$ and $-ck_z$.

It is worth noting here that the analytic properties of the
scattering matrix (or the Jost function) in the Casimir
studies are different in comparison with the standard
theory of potential scattering~\cite{Newton}. In fact they
are close to those for the Klein-Gordon equation, the role
of mass squared in Casimir calculations being played by $k_z^2$.
This implies in particular that the analytical properties of
the scattering matrix in the Casimir calculations should be
revealed by a direct analysis of its explicit form without referring to
the nonrelativistic  potential scattering.

Taking into account all this  we obtain the following
formula for the spectral zeta function in the problem under
consideration:
\begin{equation} \label{3-5}
\fl \zeta_{\mathrm{shell}}(s)=\frac{1}{\sqrt{\pi}\,
a\Gamma(s)\Gamma \left(\displaystyle\frac{3}{2}-s\right)}
\left(\frac{c}{a}\right)^{-2s}\sump\int_0^{\infty} \rmd y\,
y^{1-2s}\frac{\rmd}{\rmd y}\ln[1-\mu_{np}^2(y)]\,{,}
\end{equation}
where
\begin{equation} \label{3-6}
\mu_{np}(y)=y\left [I_{np}(y)K_{np}(y)\right ]'{.}
\end{equation}
 It is worth comparing  (\ref{3-5}) with (3.11) in
\cite{LNB}. Further we will follow the well elaborated
procedure for construction of the analytical continuation
of the formulae like (\ref{3-5}) into the left semi-plane
of the complex variable $s$ (see, for example, \cite{RNC}).
This regular procedure substantially uses the uniform
asymptotic expansion (UAE) of the modified Bessel
functions. Remarkably, the result obtained in this way can
be of arbitrary required accuracy. To obtain it, one should
keep in UAE sufficiently many terms. As far as we know,
there are no other methods to construct the analytical
continuation needed.

Upon the change of the integration variable in
(\ref{3-5}) $y=npz,\; n=1,2 \ldots $ we can use here UAE.
We content ourselves with the first two terms in UAE
\[
\ln\left\{1-\left[z\frac{\rmd}{\rmd z}(I_{np}(npz)K_{np}(npz))\right]^2\right\}
\]
\be \label{3-7}
\fl =-\frac{z^4t^6}{4n^2p^2}\left[1+\frac{t^2}{4n^2p^2}\left(3-
30t^2+35t^4+\frac{1}{2}\,z^4t^4
\right)+\Or (n^{-4})\right],\quad t=(1+z^2)^{-1/2}{.} \ee Now
we substitute (\ref{3-7}) into all the terms in (\ref{3-5})
with $n\neq 0$. The term with $n=0$ in this sum will be
treated by subtracting and adding to the logarithmic
function the quantity \be \label{3-8}
-\frac{1}{4}\frac{y^4}{(1+y^2)^3}\,{.} \ee
As a result, the
zeta function $\zeta_{\mathrm{shell}}(s)$ can be presented
now as the sum of three terms
\be \label{3-9}
\zeta_{\mathrm{shell}}(s)=Z_1(s)+Z_2(s)+Z_3(s)\,{,}
\ee
where
\be
\label{3-10}
\fl Z_1(s)=\frac{2s-1}{2\sqrt{\pi}\,a
\Gamma(s)
\Gamma\left(\displaystyle\frac{3}{2}-s\right)}
\left(\frac{c}{a}\right)^{-2s}\int_0^{\infty}\rmd y\,y^{-2s}
\left\{\ln [1-\mu_0^2(y)]+
\frac{1}{4}\,y^4t^6(y)\right\}\,{,}
\end{equation}
\be\label{3-11}
\fl Z_2(s)
=\left(\frac{c}{a}\right)^{-2s}\frac{(1-2s)(3-2s)}{64\sqrt{\pi}\,
a}\,[2p^{-1-2s}\zeta_{\mathrm{R}} (2s+1)+1]
\displaystyle\frac{\Gamma\left({\displaystyle\frac{1}{2}}+s\right)}
{\Gamma\left({\displaystyle s}\right)}\,{,}
\end{equation}
\be
\label{3-12}
\fl Z_3(s)=\left(\frac{c}{a}\right)^{-2s}
\frac{(1-2s)(3-2s)(284 s^2-104 s-235)}{61440\sqrt{\pi}a
p^{3+2s}}\,\frac{ \displaystyle
\Gamma\left(\frac{3}{2}+s\right)}{\displaystyle
\Gamma(s)} \zeta_{\mathrm{R}}(3+2s)\,{.}
\ee
As  was shown in \cite{LNB}, $Z_1(s)$ is an analytic
function of the complex variable $s$ in the region
$-3/2<\mathrm{Re}~s<1/2$. In  (\ref{3-11}) and (\ref{3-12})
$\zeta_{\mathrm{R}}(s)$ is the Riemann zeta function which accomplishes the analytical continuation of the
sum $\sum_{n=1}^{\infty}n^{-2s}$ to the left semi-plane of~$s$.

It is left now to take the limit $s\to -1/2$ in  (\ref{3-9}).
Contribution $Z_1(s)$ (\ref{3-10}) to the zeta function  does
not contain the parameter $p$, i.e.\  it remains the same as in the case of
Minkowski space-time. Hence for $Z_1(-1/2)$ the value from  \cite{LNB} can be taken
\be
\label{3-13}
Z_1(-1/2)=\frac{c}{2\pi a^2}(-0{.}6517)\,{.}
\ee
A special care should be paid when calculating the limit $s\to -1/2$
in $Z_2(s)$ (\ref{3-11}) in view of the poles of the function $\Gamma(s+1/2)$
at this point. Using the values
\be\label{3-13a}
\zeta_{\mathrm{R}} (0)=-\frac{1}{2}, \quad \zeta_{\mathrm{R}}^{\prime}(0)=-\frac{1}{2}\ln 2\pi\,{,}
\quad \Gamma(x)= \frac{1}{x}- \gamma + {\Or}(x)\,{,}
\ee
we  derive \cite{LNB,NP}
$$
\lim_{s\to -1/2}
[2p^{-1-2s}\zeta_{\mathrm{R}} (1+2s)+1]\,\Gamma\left(\frac{1}{2}+s\right)=
$$
$$
=\lim_{s\to -1/2}\left \{ 2[1-(1+2s)\ln p+\ldots][\zeta_{\mathrm{R}} (0)
+\zeta_{\mathrm{R}}^{\prime}(0)(1+2s)+
\ldots]+1\right \}\left( \frac{2}{1+2s} -
\gamma+\ldots\right)
$$
\be\label{3-14}
=-2\ln \frac{2\pi}{p} \,{.}
\ee
Thus  we have for $Z_2(-1/2)$
\be
\label{3-15}
Z_2(-1/2)=\frac{c}{8\pi a^2}\ln{\frac{2\pi}{p}}\,{.}
\ee
The third term  (\ref{3-12}) gives
\be
\label{3-16}
Z_3(-1/2)=\frac{c}{2\pi a^2}\frac{7}{480}\frac{\pi^2}{6p^2}\,{.}
\end{equation}

Gathering together $Z_i(-1/2),\;i=1,2,3$ we obtain the finite value
for the vacuum energy of electromagnetic field in the problem
under consideration: \be \label{3-17}
E_{\mathrm{shell}}=\frac{\hbar}{2}\,\zeta_{\mathrm{shell}}\left
(-\,\frac{1}{2} \right ){,} \ee where \be \label{3-18}
\zeta_{\mathrm{shell}}\left (-\frac{1}{2} \right )=\frac{c}{2\pi
a^2}\left( -0{.}6517 +\frac{1}{4} \ln\frac{2\pi}{p} +
\frac{7}{480}\frac{\pi^2}{6}\frac{1}{p^2} \right ){.} \ee At
$p=1$ we reproduce (3.35) from \cite{LNB}.

The consideration
presented here  can be extended to the next terms
in the UAE of the Bessel functions (\ref{3-7}) in a
straightforward way. Therefore we shall not present here
these rather cumbersome expressions.
\subsection{Diaphanous  matching conditions}
\label{iso-vacuum}
In this case, we have continuous spectrum of electromagnetic
oscillations (\ref{2-10a}). As a result only the second term
in (\ref{3-4}) survives. When going to the imaginary
frequencies the contours $C^\pm_{\Omega}$ should be used
(see figure 1b). We again obtain the integral representation
(\ref{3-5}) for the spectral zeta function, but now $\mu_{np}
(y)$ should be multiplied by the parameter $\xi^2$:
 \be \label{3-19}
\xi^2=\frac{(\varepsilon_1-\varepsilon_2)^2}{(\varepsilon_1+\varepsilon_2)^2}
=\frac{(\mu_1-\mu_2)^2}{(\mu_1+\mu_2)^2}\leq 1{.} \ee Performing
in the same way as for the perfectly conducting boundary
conditions, we arrive at the result \be \label{3-20}
\zeta^{\mathrm{lin}}_{\mathrm{cyl}}\left (-\frac{1}{2} \right )=\frac{c\xi^2}{2\pi
a^2}\left
(-0.490878+\frac{1}{4}\ln{\frac{2\pi}{p}}+\frac{\pi^2}{288}
-\frac{\pi^2}{360}\frac{7}{1920}\frac{1}{p^4} \right){.}
\ee
We are considering only the contribution to the zeta
function which is linear in $\xi^2$. When $p\to 1$
the derived value for
$\zeta^{\mathrm{lin}}_{\mathrm{cyl}}(-1/2)$ vanishes \cite{MNN,NP-PRD,Klich-Romeo}.
As was noted  earlier, in cosmic string physics the parameter $p$ is close to 1.
Hence, the vacuum energy of electromagnetic field for these boundary conditions
\be
\label{3-21}
E=\frac{\hbar}{2}\,\zeta^{\mathrm{lin}}_{\mathrm{cyl}}\left (-\, \frac{1}{2}\right )
\ee
will be also small.
\subsection{Semitransparent matching conditions}
At first, we assume that $g>0$ and, as a consequence, the
spectrum of excitations of the field $\varphi (t,
\mathbf{x})$ is pure continuous (see section \ref{2b}).
When constructing the spectral zeta function, we carry out
the integration over the continuous spectrum (\ref{2-10a})
by making use of (\ref{3-4}) keeping there the second
integral term alone. Further there is no need to subtract
here the contribution of Minkowski space-time (the term
obtained in the limit $a\to \infty $). The point is that such a
subtraction has been already done  in  (\ref{3-3}). By
making use of the exact expression for the $S$-matrix (\ref{2-30}) and the
integration contours shown in figure (1b),
we obtain the following representation for the zeta
function
\begin{equation} \label{3-23}
\zeta_{\mathrm{s-t}}(s)=C(s)\sump\int_0^{\infty} \rmd y\,
y^{1-2s}\frac{\rmd}{\rmd y}\ln \left [1+gI_{\nu}(y)K_{\nu}(y) \right ]\,{,}
\end{equation}
where $\nu=np$ and
\begin{equation} \label{3-24}
C(s)=\frac{c^{-2s}}{a^{1-2s}\sqrt{\pi }\, \Gamma (s) \Gamma (3/2-s)}\,{,} \quad
C(-1/2)=-\frac{c}{2\pi a^2}
\,{.}
\end{equation}

Further we shall consider the zeta function (\ref{3-23}) in
the linear  approximation with respect to $g$
\begin{equation} \label{3-25}
\zeta_{\mathrm{s-t}}(s)=g\,C(s)\sump\int_0^{\infty} \rmd y\,
y^{1-2s}\frac{\rmd}{\rmd y}\left [I_{\nu}(y)K_{\nu}(y)\right ]\,{.}
\end{equation}
Analytical continuation of this function to the left
semi-plane of the complex variable $s$ will be accomplished
in the standard way (see, for example, \cite{RNC}): we add
and subtract to the integrand in (\ref{3-25}) for $n \geq
1$ a few first terms of the UAE for the product of the
modified Bessel functions $I_\nu(\nu z) K_\nu(\nu z)$:
\begin{equation} \label{3-25a}
I_\nu (\nu z) K_\nu(\nu z)\simeq
\sum_{k=0}^\infty \frac{P_k(t)}{\nu^{2k+1}}\,{,} \quad t=\frac{1}{\sqrt{1+z^2}} \,{,}
\end{equation}
where
\numparts
\label{alleqs}
\begin{eqnarray}
P_0(t)&=&\frac{t}{2}{,}\label{eqa} \\
P_1(t)&=&\frac{t^3}{16}\,(1-6t^2+5t^4)\label{eqb}{,} \\
P_2(t)&=&\frac{t^5}{256}\,(27-580t^2+2170t^4-2772t^6+1155t^8)\label{eqc}{,}\\
P_3(t)&=&\frac{t^7}{2048}(t^2-1)(425425 t^{10}   - 1106105 t^8  + 1014442 t^6  \\
        && - 383570 t^4 + 51445 t^2  - 1125)\,{,}
\end{eqnarray}
\endnumparts
and so on~\cite{AS}.

Do not pretending
to high accuracy,  we take in UAE (\ref{3-25a}) for subtracting  and adding
only two terms
\begin{equation} \label{3-26}
I_\nu (\nu z) K_\nu(\nu z)\simeq \sum_{k=0}^1\frac{P_k(t)}{\nu^{2k+1}}=\frac{t}{2\nu}\left [
1+ \frac{t^2(1-6 t^2 +5 t^4)}{8 \nu^2}+\ldots
\right ]{.}
\end{equation}
In the case of the product $I_0(y)K_0(y)$ we add and
subtract the usual asymptotic of this expression  when $y\to \infty$
\begin{equation} \label{3-27}
\frac{1}{2\sqrt{1+2y^2}}\,{.}
\end{equation}

As a result, the zeta function (\ref{3-25}) can be
represented now as the sum of four terms
\begin{equation} \label{3-28}
\zeta_{\mathrm{s-t}}(s)=g\,C(s)\sum_{i=1}^4 Z_i(s)\,{,}
\end{equation}
where
\begin{equation} \label{3-29}
\fl Z_1(s)=\frac{1}{2}\int_0^\infty \rmd y\, y^{1-2s}\frac{\rmd}{\rmd y}\left [
I_0(y)K_0(y)- \frac{1}{2\sqrt{1+2y^2}}\right ]{,}
\end{equation}
\begin{equation} \label{3-30}
\fl Z_2(s)=\frac{1}{2}\left (
\sum_{n=1}^\infty \nu^{-2s}+\frac{1}{2}
\right )\int_0^\infty \rmd z\, z^{1-2s}\frac{\rmd}{\rmd z}\,t\,{,}
\end{equation}
\begin{equation} \label{3-31}
\fl Z_3(s)=\frac{1}{16}\sum_{n=1}^\infty \nu^{-2-2s}
\int_0^\infty \rmd z\, z^{1-2s}\frac{\rmd}{\rmd z}\,t^3(1-6t^2+5t^4)\,{,}
\end{equation}
\begin{equation} \label{3-32}
\fl Z_4(s)=\sum_{n=1}^\infty \nu^{1-2s}
\int_0^\infty \rmd z\, z^{1-2s}\frac{\rmd}{\rmd z}\left \{
I_\nu(\nu z)K_\nu(\nu z)-
\frac{t}{2\nu}\left [
1+ \frac{t^2(1-6 t^2 +5 t^4)}{8 \nu^2} \right ]
\right \}{.}
\end{equation}

The obtained formulae (see  (\ref{3-29}), (\ref{3-30}), and
(\ref{3-31})) show clear the basic idea of making use of
the UAE for analytic continuation: this method allows one
to factorize the divergences which are originated in the
sum over $n$ and in integration over $y$ or $z$. In what
follows we express the result of the summation in terms of
the Riemannian zeta function and the results of integration
through the gamma function. It is this that provides us
with explicit analytic continuation needed. The last
term, $Z_4(s)$, can be evaluated only numerically upon
fixing the parameter $p\quad (\nu=n p$) and the value of
the variable $s$, for example, $s=-1/2$. As a rule, this
term gives a small correction \cite{MNN,NP}. Fortunately,
in the problem under consideration this term seems do not
contribute to the Casimir energy for arbitrary values of
$p$ (see below).

Taking into account the behaviour of $(\rmd/\rmd y)[I_0(y)K_0(y)]$
at the origin and at infinity
\begin{equation}
\label{3-33}
\frac{\rmd}{\rmd y}\left [I_0(y) K_0(y)\right ]\simeq
\cases{\displaystyle-\frac{1}{y},\quad y\to 0, \cr
\displaystyle -\frac{1}{2y^2} -  \frac{3}{16 y^4}\,{,}\quad  y\to \infty\,{,}\cr}
\end{equation}
we infer, that the function $Z_1(s)$ defined by
(\ref{3-29}) is an analytic function in the region
\begin{equation}
\label{3-34}
-1 < \mathrm{Re}\;s<1/2\,{.}
\end{equation}
The upper bound in (\ref{3-34}), which is caused by the
behaviour of the integrand in (\ref{3-29}), can be removed by
introducing the infrared cutoff or the mass of the field
$\varphi (t,\mathbf{x})$ (see, for example, \cite[p.~4524]{NP}).
At the end of calculations this cutoff (or mass) should be set
zero. At the point $s=-1/2$ we obtain numerically
\begin{equation}
\label{3-35}
g\, C(-1/2) Z_1(-1/2)=\frac{cg}{4\pi a^2}\cdot 0{.}999\ldots\,{.}
\end{equation}

Integral in  (\ref{3-30}) exists originally in the region
\begin{equation}
\label{3-36}
0 < \mathrm{Re}\;s<3/2\,{,}
\end{equation}
and the sum over $n$ in this formula is finite in the domain
\begin{equation}
\label{3-37}
\mathrm{Re}\;s>1/2\,{.}
\end{equation}
Thus, in the region
\begin{equation}
\label{3-38}
1/2< \mathrm{Re}\;s<3/2
\end{equation}
Equation (\ref{3-30}) defines an analytic function of the
variable $s$. We can analytically continue this function to
the whole complex plane $s$, except for simple poles at
some isolated points. For this purpose we have to put
\begin{equation}
\label{3-39}
\sum_{n=1}^\infty \nu^{-2s}=p^{-2s}\zeta_{\mathrm{R}}(2s)
\,{,}
\end{equation}
\begin{equation}
\label{3-40}
\int_0^\infty \rmd z\,z^{1-2s}\frac{\rmd}{\rmd z}\,t=-\int_0^\infty \rmd z\,z^{2-2s}t^3=-\frac{1}{2}\,
\frac{\Gamma(s)\Gamma(3/2-s)}{\Gamma(3/2)}
\,{.}
\end{equation}
We have used here the table integral~\cite{GR}
\begin{equation}
\label{3-41}
\int_0^\infty z^{\alpha-1}t^\beta \rmd z=\frac{1}{2}\,\frac{\displaystyle \Gamma\left (
\frac{\alpha  +\beta}{2}\right )\Gamma\left (
-\frac{\alpha }{2}\right )}{\displaystyle \Gamma\left (
\frac{\beta }{2}\right )}\,{.}
\end{equation}
The right-hand side of (\ref{3-41}) has no singularities (poles) in the region
\begin{equation}
\label{3-42}
\mathrm{Re}\,\alpha <0,\quad \mathrm{Re}\,(\alpha+\beta) >0\,{.}
\end{equation}
Admitting the appearance of poles, we define the left-hand side of (\ref{3-40})
in the whole complex plane $s$ by the right-hand side in this formula.

Finally we have
\[ \fl gC(s)Z_2(s)=-\frac{g}{8}  C(s)\left [
2p^{-2s}\zeta _{\mathrm{R}}(2s)+1
\right ]\frac{\displaystyle\Gamma (s)\Gamma\left
 ({3}/{2}-s \right )}{\displaystyle \Gamma\left ({3}/{2}\right )}\]
\begin{equation}
\label{3-43}
=
-\frac{gc^{-2s}}{2\pi a^{1-2s}}\left [
p^{-2s}\zeta_{\mathrm{R}}(2s)+\frac{1}{2}
\right]{.}
\end{equation}
At the point $s=-1/2$ it gives
\begin{equation}
\label{3-44}
g\,C(-1/2)Z_2(-1/2)=-\frac{c g}{2\pi a^2}\left [
p\,\zeta _{\mathrm{R}}(-1)+\frac{1}{2}
\right ]
= -\frac{c g}{4\pi a^2}\left (
1-\frac{p}{6}
\right )
{.}
\end{equation}

In the same way
we can construct the analytic continuation of the function $Z_3(s)$, which is
defined originally by (\ref{3-31}) in the region
\begin{equation}
\label{3-45}
0<\mathrm{Re}\, s<3/2\,{.}
\end{equation}
In the whole complex plane $s$, we have
\begin{equation}
\label{3-46}
g\,C(s)Z_3(s)=-\frac{g\,c^{-2s}p^{-2-2s}}{24\pi a^{1-2s}}\zeta _{\mathrm{R}}(2+2s)(2s+1)(2s-1)\,{.}
\end{equation}
At the point $s=-1/2$ we find
\begin{equation}
\label{3-47}
g\,C(-1/2)Z_3(-1/2)=-\frac{c\,g}{24 \pi a^2 p}
\,{.}
\end{equation}

The functions $Z_i(s),\;i=1,2,3$ are defined by initial formulae (\ref{3-29}) --
(\ref{3-31}) in the domains of complex plane $s$ which have a common strip, namely
\begin{equation}
\label{3-16a}
1/2<\mathrm{Re}\,s<3/2
\end{equation}
(see  (\ref{3-34}) with the note following,
(\ref{3-38}),  and (\ref{3-45})). This implies, that the
analytic continuations of individual functions
$Z_i(s)\;i=1,2,3$, constructed above, applies also to the sum
$\sum_{i=1}^3Z_i(s)$.

Gathering together the contributions (\ref{3-35}), (\ref{3-44}), and (\ref{3-47})
we get
\begin{equation}
\label{3-47w}
\zeta_{\mathrm{s-t}}(-1/2)=\frac{cg}{24\pi a^2}\left (
p-\frac{1}{p}
\right ){.}
\end{equation}

It seems that this answer is exact (in the linear in $g$
approximation). The point is the last term, $Z_4(s)$ in the sum
(\ref{3-28}) does not contribute to the vacuum energy for
arbitrary $p$. This assertion is based on the following fact
noted, for the first time, in the paper \cite[Appendix A]{CPMK}:
Upon  substituting in (\ref{3-32}) the product $I_{\nu}(\nu
z)K_{\nu}(\nu z)$ by the uniform asymptotic expansion (\ref{3-25a})
and setting here  $s=-1/2$, we obtain
\[
\fl Z_4(-1/2)=\sum_{k=2}^\infty \int_0^\infty \rmd z\, z^2 \frac{\rmd}{\rmd z}P_k(t)
\sum_{n=1}^\infty
\frac{1}{(np)^{2k-1}}
\]
\begin{equation}
\label{3-47a}
=\sum_{k=2}^\infty p^{1-2k}\zeta_{\mathrm{R}}(2k-1)
\int_0^\infty \rmd z\, z^2 \frac{\rmd}{\rmd z} P_k(t)\,{.}
\end{equation}
Analytical integration in (\ref{3-47a}) gives zero
\begin{equation}
\label{3-47b}
\int_0^\infty \rmd z\, z^2 \frac{\rmd}{\rmd z}P_k(t)=0\,{.}
\end{equation}
We have checked this relation for $k=2,3$ by making use of the table integral~\cite{GR}
\begin{equation}\label{3-47c}
\fl \int\limits_0^{\infty}\rmd z \,z^{1-s}\frac{\rmd}{\rmd z}t^{2(\rho-1)}=
(1-\rho)\,\frac{\Gamma\left(\frac{\displaystyle 3-s}{\displaystyle
 2}\right)
\Gamma\left(\rho-\frac{\displaystyle  3-s}{\displaystyle  2}
\right)}{\Gamma(\rho)}, \quad 3-2 \,\mathrm{ Re }\,\rho<\mathrm{Re}\,s<3.
\end{equation}

Certainly, it is important  to bring to light the origin of these null results (see also
\cite{Milton,CPMK}).

Proceeding from all this, we conclude that the exact value
(in the linear in $g$ approximation) of the  Casimir energy in the
problem at hand is
\begin{equation}
\label{3-48}
E_{\mathrm{s-t}} =\frac{\hbar}{2}\,
\zeta_{\mathrm{s-t}}(-1/2)=
\frac{c\hbar g}{48\pi a^2}\left (
p-\frac{1}{p}
\right ){,} \quad g>0\,{.}
\end{equation}
The vacuum energy (\ref{3-48}) is positive because we have
assumed that the constant $g$ is positive. In the limit $p=1$
 (\ref{3-48}) reproduces the vanishing vacuum energy of the
cylindric  delta-potential \cite{Milton,CPMK}.
\noindent
\begin{figure}[th]
\noindent \centerline{
\includegraphics[width=7.20cm]{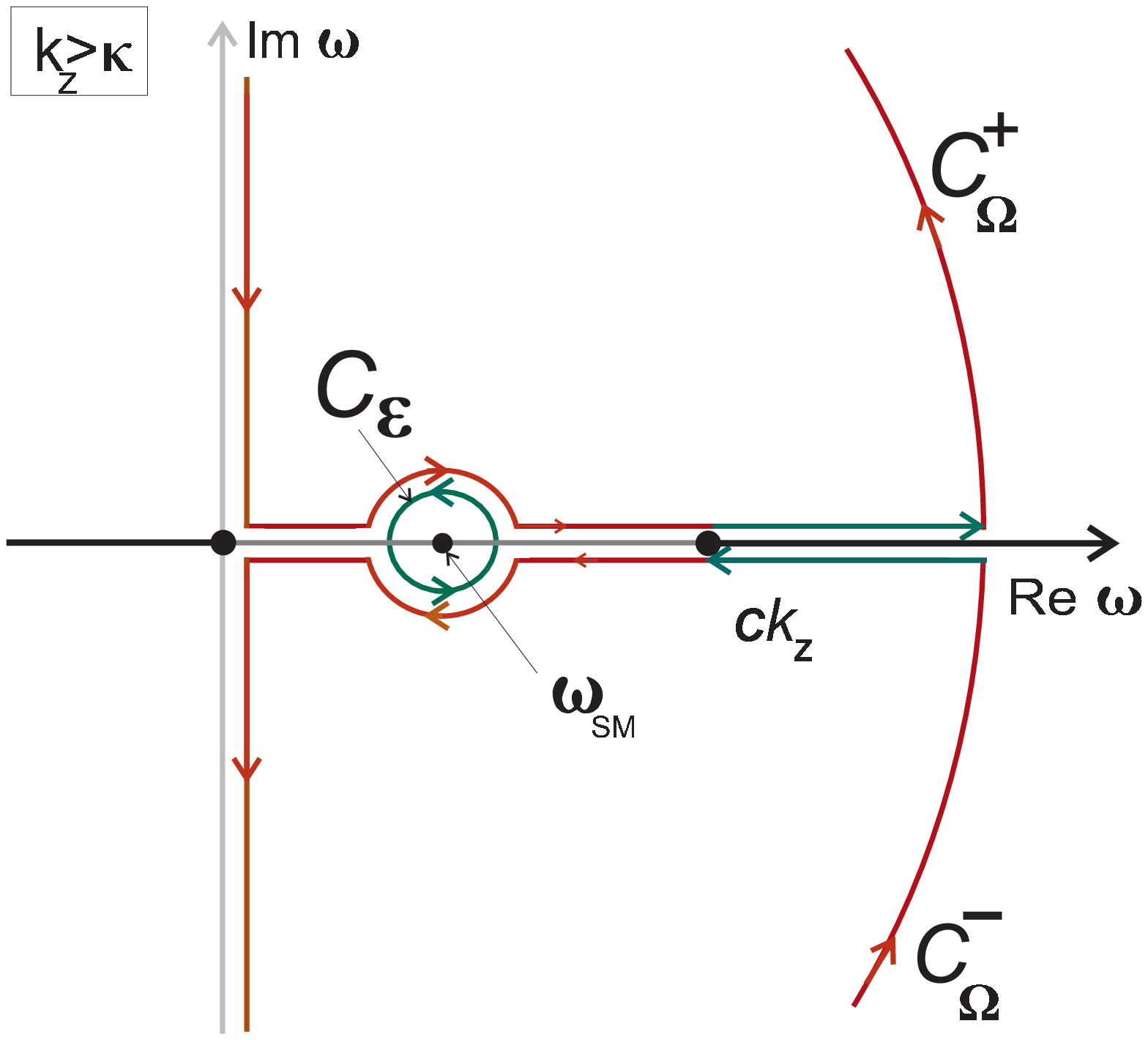}
\hspace{20mm}
\includegraphics[width=6.7cm]{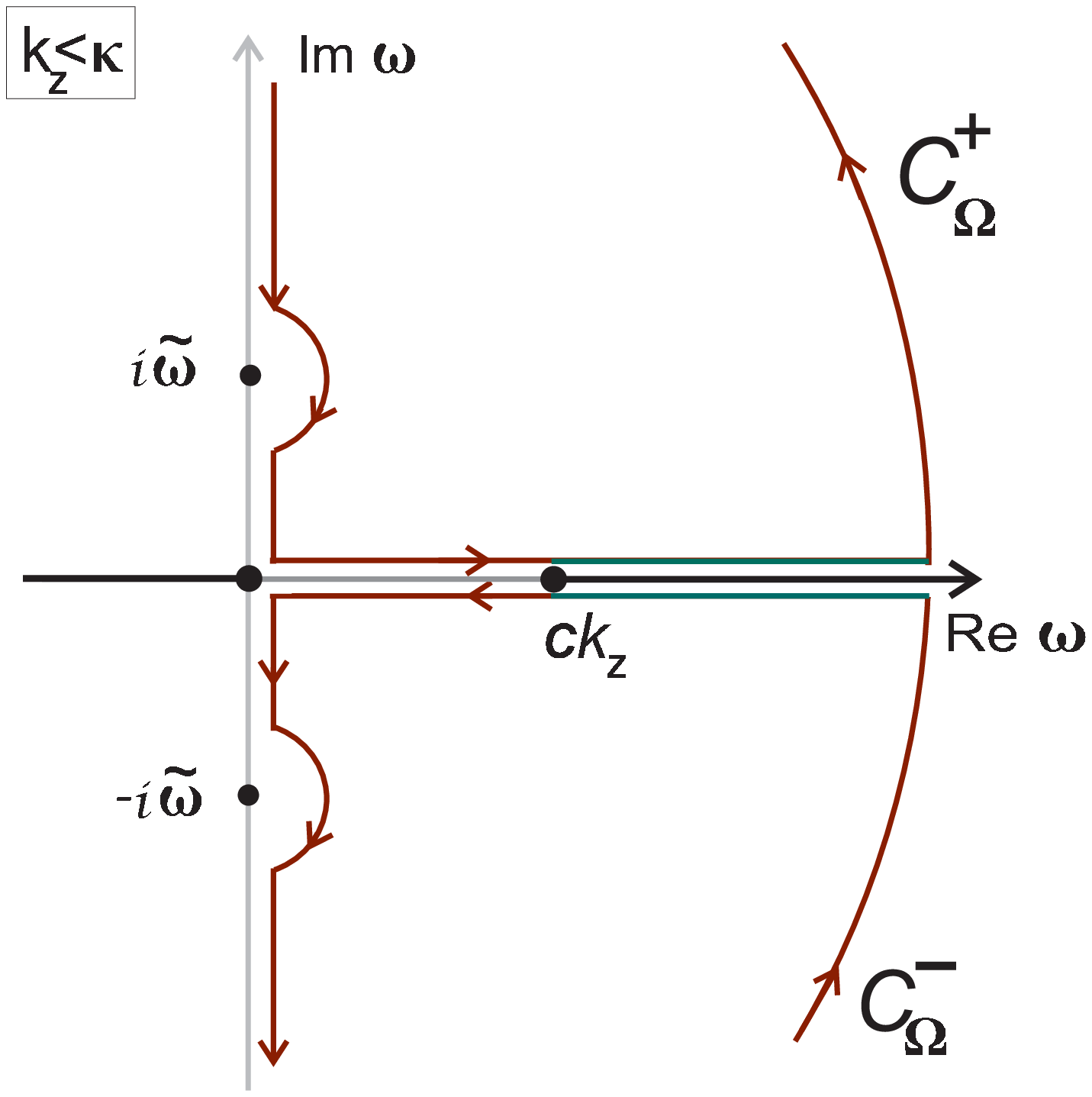}
} \caption{Contours in the complex $\omega$ plane which should be
used when going on to the integration over imaginary frequencies in
the case of spectrum having continuous branch and surface modes.
There are two cuts along the real axis: $-\infty < \omega < 0$ and
$ck_z< \omega < \infty$.}
 \label{fig2}
\end{figure}

Now we address the consideration of negative constant  $g$, when
the spectrum of the field $\varphi (t,\mathbf{x})$ has a
continuous part (\ref{2-10a}) and surface modes with frequencies
(\ref{2-47}). In conformity with this we represent the
zeta-regularized vacuum energy $E(s)$ as the sum of two terms
\begin{equation}
\label{3-55a}
E(s)=E_{\mathrm{sm}}(s)+E_{\mathrm{cont}}(s)
\end{equation}
(see  (\ref{3-4})). Transition to the imaginary frequencies
in $E_{\mathrm{cont}}(s)$ is accomplished by making use of the
integration along the contours $C^+_\Omega$ and $C^-_\Omega$
shown in figure 2. One can see from this figure, that
$E_{\mathrm{cont}}(s)$ involves, with an opposite sign, the
contributions of the surface modes $\omega_{\mathrm{sm}}$
((\ref{2-47}) with $|k_z|>\kappa$) and the contribution of pure
imaginary frequencies $\pm i \tilde \omega$ (see
(\ref{2-49})). The contribution of $\omega_{\mathrm{sm}}$ to
$E_{\mathrm{cont}}(s)$ is canceled with $E_{\mathrm{sm}}(s)$
(integration along the contour $C_{\varepsilon}$, the
contributions of the imaginary frequencies $\pm i\tilde \omega $
are mutually canceled at $s=-1/2$.

 As a result, we arrive
again at the  formula (\ref{3-23}) with an alone change, namely,
the integral over imaginary frequencies\footnote{The variable
$y$ is not equal exactly to $\rmi\,\omega$, more precisely, it is
related to $i\lambda$ in (\ref{2-6}).}  $y$ should be
treated as the principal value integral
\begin{equation} \label{3-49}
E(s)=\frac{1}{2}C(s)\sump\mathrm{P.V.}\int_0^{\infty} \rmd y\,
y^{1-2s}\frac{\rmd}{\rmd y}\ln \left [1-|g|I_{\nu}(y)K_{\nu}(y) \right ]\,{,}
\end{equation}
where $C(s)$ is defined, as before, in (\ref{3-24}). Obviously, in the
linear in $-|g|$ approximation we do  not `feel' the P.V.-prescription.
Hence in this approximation we have the result (\ref{3-48}) with the negative sign:
\begin{equation}
\label{3-50}
E_{\mathrm{s-t}} =-\,\frac{c\hbar |g|}{48\pi a^2}\left (
p-\frac{1}{p}
\right ){,} \quad g<0\,{.}
\end{equation}

In \cite{Milton,CPMK} dealing
with a massless scalar field in the background of
semitransparent cylindrical shell ($p=1$), the negative values of
$g$ were not considered at all with the aim `to avoid the
appearance of negative eigenfrequencies'~\cite[p.\ 29]{Milton}.
The vacuum energy of massive scalar field in the background of a
cylindrical $\delta$-potential was considered in
\cite{Scandurra,Scandurra-thesis} for both repulsive and
attractive potentials. The possibility of bound states were not
investigated there and it was noted that for $g<0$ the
renormalized Casimir energy acquires complex values.

\section{High temperature expansions}
Quantum field theory predicts
a specific temperature dependence for the
energy of radiation connected  with a heated
body. One cannot anticipate here the Planck spectrum and the Boltzmann law.
These  laws hold for the photons
being in unbounded space and  nevertheless  possessing nonzero
temperature.  For such photons a notion of black body radiation has
been introduced as far back as in pre-quantum physics time. It is
clear that  in the general case,  the spectrum of a heat radiation
and its temperature behaviour  should depend
on the shape of a heated body. By the way this dependence is
important for many radio engineering equipments (thermal noise of
antennae, waveguides and so on).

For obtaining the asymptotic behaviour of the thermodynamical
functions, a powerful method of the zeta function technique and
the heat kernel expansion \cite{BNP} can be used. It is
important that for these calculations it is sufficient to find
the zeta function and the heat kernel coefficients not for a
complete operator, determining the quadratic action of the
quantum field under consideration, but only for the space part
of this operator, i.e.\ one has to calculate first  the zeta
function that is used for the Casimir calculations at zero
temperature. This is an essential merit of this approach. It is
these spectral zeta functions that have been constructed in the
preceding sections. Obviously, these results can be used for
obtaining the high-temperature expansions for the thermodynamic
potentials in the problems under consideration.

The high-temperature asymptotics for the Helmholtz free energy
has the form \cite{RNC,BNP}
\begin{eqnarray}
\fl F(T) \simeq -\frac{T}{2}\zeta'(0)+B_0\frac{T^4}{\hbar^3}\,
\frac{\pi^2}{90}-B_{1/2}\,\frac{T^3}{4\pi^{3/2} \hbar^2 }
\zeta_{\mathrm {R}}(3) -\frac{B_1}{24}\frac{T^2}{\hbar}
+\frac{B_{3/2}}{(4\pi)^{3/2}}\,T\,\ln\frac{\hbar}{T}\nonumber\\
-\frac{B_2}{16\pi^2}\,\hbar\, \left[\ln\left(\frac{\hbar}{4 \pi
T}\right)+\gamma\right]-\frac{B_{5/2}}{(4\pi)^{3/2}}\frac{\hbar^2}{24
T}\nonumber\\ -T\sum_{n\geq
3}\frac{B_n}{(4\pi)^{3/2}}\left(\frac{\hbar}{2 \pi
T}\right)^{2n-3}\,\Gamma(n-3/2)\,\zeta_{\mathrm{R}}(2 n-3)\,{,}
\quad T\to \infty {.} \label{4-1}
\end{eqnarray}
Here $\gamma$ is the Euler constant and $B_{n/2},\;
n=0,1,2,\ldots$ are the heat kernel coefficients which can be
calculated by making use of the spectral zeta function in the
problem at hand
\begin{equation}
\label{4-2} \frac{B_{n/2}}{(4 \pi)^{3/2}}=
\lim_{s\to\frac{3-n}{2}} \left ( s+\frac{n-3}{2} \right )\Gamma
(s) \zeta(s),\quad n=0,1,2, \ldots \,{.}
\end{equation}
In order to use this formula the  function $\zeta (s)$
should be known in the vicinity of the following points
\begin{equation}\label{4-3}
s=\frac{3}{2},\, 1,\, \frac{1}{2},\, 0, \, \ldots \,{.}
 \end{equation}
The heat kernel coefficient $B_{n/2}$ is different from zero only when the product
$\Gamma (s)\zeta(s)$ has a (simple) pole at the point $s=(3-n)/2,\;n=0,1,2, \ldots $ .

The asymptotic expansions for the internal energy $U(T)$ and the
entropy $S(T)$ are deduced from    (\ref{4-1}) through  the
thermodynamic relations
\begin{eqnarray}
U(T)&=&-T^2\frac{\partial}{\partial T}\left(T^{-1}F(T)\right),
\label{4-13}\\ S(T)&=&T^{-1}\left(U(T)-F(T)\right)=
 - \frac{\partial F}{\partial T}{.}
\label{4-14}
\end{eqnarray}

For compact manifolds  $B_0\sim V$ and $B_{1/2}\sim S$, where $V$ is
the volume of the manifold and $S$ is the area of its boundary.
In view of this the second term in  the right-hand side of
(\ref{4-1}) gives the Stefan-Boltzmann law ($U(T)\sim VT^4$).

In the case of noncompact manifolds the {\it relative} spectral
functions (zeta function and heat kernel) are employed  (see,
for example, \cite{Mueller,Hurt}). When  defining these functions the
contribution due to  free space is subtracted. We have noted
earlier that the scattering formalism implements this automatically.
As a result, in our approach $B_0/V\to 0$, when $V\to \infty$
and $B_{1/2}/S\to 0$, when $S\to \infty$ (see for details \cite[p.\ 446]{NPD}).

Summarizing, we may infer, that  the approach applied gives in fact
the corrections to the Stefan-Boltzmann law
which are determined by the specific spectrum of electromagnetic
oscillations (or other fields), i.e.\ by the geometry of the problem under
consideration. It should be noted that these corrections are not
proportional to the volume $V$ of the system, hence we cannot
say about the corrections to the energy density of
electromagnetic energy. The area of practical employment of
these  thermodynamical asymtotics is calculation of the
temperature dependence of respective Casimir forces.

The heat kernel coefficients needed in (\ref{4-1}) were
calculated in our previous article \cite{NPD} and in Appendix~A
and Appendix~B of the present paper. The value of $\zeta'(0)$
can be found by technique employed for this purpose, for
example, in  \cite{BNP}. By making use of all this we
obtain the thermodynamical asymptotics in the geometry under consideration.

For electromagnetic field with perfectly conducting boundary
conditions these asymptotics are
\begin{eqnarray}
F(T)&\simeq&-\frac{T}{2}\zeta'(0)+\frac{3}{64 p}\frac{T}{a}\ln\frac{\hbar}{T}-
\frac{153}{2048}\frac{1}{4\pi}\frac{\hbar^3c^2}{24 T a^3},\\
U(T)&\simeq &\frac{3}{16}\;\frac{T}{4 \pi a}-\frac{153}{2048}\;\frac{1}{4\pi}
\;\frac{\hbar^2c^2}{12 T a^3}
,\\
S(T)&\simeq&\frac{1}{2}\zeta'(0)+\frac{3}{64 p T}\left (
1-\ln\frac{\hbar}{T}\right )-\frac{153}{2048}\;\frac{1}{4\pi}\;\frac{\hbar^2c^2}{24 T^2 a^3}\,{,}
\end{eqnarray}
where
\begin{equation}
\zeta'(0)=\frac{1}{a}\left[0.53490+\frac{1}{32\,p}\left (3\ln\frac{
a}{2pc}+3\gamma-4-\frac{47}{256 p^2}\zeta_{\mathrm{R}}(3)\right )\right]{.}
\end{equation}
For isorefractive matching conditions we have
\begin{eqnarray}
\fl F(T)&\simeq&-\frac{T}{2}\zeta'(0)+\frac{2}{45}\frac{T^4}{\hbar^3}
\frac{a^2\xi^2\pi^3}{c^3}+
\frac{3}{16}\frac{\xi^2}{4 pa}T\ln\frac{\hbar}{T}-\frac{45}{1024}
\frac{\xi^2c^2}{pa^3}\frac{\hbar^2}{96T},\\
\fl U(T)&\simeq&-\frac{T^4}{\hbar^3}\frac{2\pi^3}{15}\frac{ a^2\xi^2}{c^3}+
\frac{3}{64 p}\frac{\xi^2 T}{a}-\frac{45}{1024}\frac{\xi^2\hbar^2c^2}{48 p T a^3}
,\\
\fl S(T)&\simeq&\frac{1}{2}\zeta'(0)-\frac{8\pi^3}{45}\frac{a^2\xi^2}{c^3}\frac{T^3}{\hbar ^3}+
\frac{3}{16}\frac{\xi^2}{4p a}\left(1-\ln\frac{\hbar}{T}\right)
-\frac{45}{1024}\frac{\xi^2c^2}{p a^3}\frac{\hbar^2}{94 T^2}\,{,}
\end{eqnarray}
where
\begin{equation}
\zeta'(0)=\frac{\xi^2}{a}\left[0.28428+\frac{1}{32 p}\left(3\ln\frac{a}{2p c}+
3\gamma-4-\frac{27}{128p^2}\zeta_{\mathrm{R}}(3)\right)\right]{.}
\end{equation}
And finally for a massless scalar field with $\delta$-potential ($g>0) $
we obtain
\begin{eqnarray}
F(T)&\simeq&-\frac{T}{2}\zeta'(0)-\frac{T^4}{\hbar^3}\frac{\pi^3
g a^2}{90c^3}-\frac{\pi g T^2}{12\hbar c}\left ( 1-\frac{1}{p}
\right ),\\
U(T)&\simeq&\frac{T^4}{\hbar^3c^3}\frac{\pi^3}{30}g a^2+\frac{\pi g
T^2}{12\hbar c}\left (1-\frac{1}{p}
\right ),\\
S(T)&\simeq&\frac{1}{2}\zeta'(0)+\frac{T^3}{\hbar^3}\frac{2\pi^3
ga^2}{45 c^3}+\frac{\pi g T}{6\hbar c}\left (1-\frac{1}{p} \right ){,}
\end{eqnarray}
where
\begin{equation}
\zeta'(0)=\frac{g}{\pi a}\left(-0.9818+\frac{\pi^2}{144 p^2}+\frac{1}{2}\ln\frac{2\pi}{p}\right){.}
\end{equation}

\section{Conclusions}
We have demonstrated that enclosing the conical singularity
with a cylindrical surface and imposing here appropriate
boundary conditions on quantized fields render the total
vacuum energy of these fields finite, at least,  in the zeta
renormalization technique.

It is worth noting here that usually conical singularity is
considered as  the singularity of the curvature tensor
\cite{Starobinsky,Fursaev}. We have treated this
singularity as a non-smoothness of the boundary \cite{NPD}.

Calculation of the Casimir energy for configuration close
to those considered in our paper has attracted much
attention in recent studies. Ellingsen,  Brevik, and Milton
\cite{BEM,EBM,BEM-Proc,MWK,EBM-t} calculated the vacuum
energy for wedge and cylinder  geometry which includes, as
a special case, our cone configuration.  The periodic
boundary conditions in angular variable should be used for this choice.
Only this case has rigorous justification for calculations conducted
(vanishing of the heat kernel coefficient $B_2$).

The local energy density for a wedge with a coaxial cylindrical
shell was considered in papers
\cite{RS,BMBS,Sah,BMMST,BMSah}. The singular behaviour of
this quantity at $r=0$ was noted in these studies. This
singularity forbade  the calculation of the respective
total vacuum energy.

As a byproduct we have presented here a complete analysis of the
spectral problem for a scalar massless field in the background
of the $\delta$-potential located at a cylinder lateral surface, both the
$\delta$-wall and the $\delta$-well being considered.

\ack
This study has been accomplished by the  financial support
of the Russian Foundation for Basic Research (Grants No.\
09-02-12417, No.\   10-02-01304, and No.\ 11-02-12232).
The authors are indebted to A.V.\ Nesterenko for help in preparing figure 1.

\appendix
\section{The heat kernel coefficients  for isorefractive boundary conditions}
Here we calculate the heat kernel coefficients for
electromagnetic field obeying  to the isorefractive matching
conditions on the cylindrical surface of radius $a$ in the
conical space. We shall use the
relation (\ref{4-2}) between the heat kernel coefficients and
the relative spectral zeta function $\zeta (s)$. As was noted in
section \ref{iso-vacuum} the spectral function in the problem at hand
is given by  (\ref{3-5}), where $\mu_{np} (y)$ should be
multiplied now by the parameter $\xi^2$ given in (\ref{3-19}).
Thus we have
\begin{equation}
\zeta(s)=C(s)\mathop{{\sum}'}_{n=0}^{\infty}\int\limits_{0}^{\infty}\rmd
y\, y^{1-2 s}\frac{\rmd}{\rmd y} \ln\left [1-\xi^2\mu_{\nu}^2(y)\right
]\,{,} \label{A1}
\end{equation}
where $\nu=np$, $\mu_\nu(y)$ is explained in  (\ref{3-6}), and
$C(s)$ is defined in (\ref{3-24}). Further we separate in (\ref{A1}) the contribution
due to the term with $n=0$ denoting the rest by  $\overline \zeta (s)$:
\begin{equation}
\zeta(s)=\zeta_0(s)+\overline \zeta (s)\,{.}
\label{A2}
\end{equation}

In the $\xi^2$-approximation $\zeta_0(s)$  reads
\begin{equation}
\label{A3}
\zeta_0(s)=-\frac{\xi^2}{2}\,C(s)\int\limits_{0}^{\infty}\rmd y\, y^{1-2 s}\frac{\rmd}{\rmd y}\mu_0^2(y).
\end{equation}
The integral over $y$ exists in the strip
\begin{equation}
\label{A3a}
-1/2<\mathrm{Re }\,s<3/2,
\end{equation}
because
\begin{equation}
\label{A4}
\fl \frac{\rmd\mu_0^2}{\rmd y}=4\left(\ln\frac{y}{2}+\gamma+\frac{1}{2}\right)y+{\Or}(y^3),\;{y\to0};
\quad \frac{\rmd\mu_0^2}{\rmd y}=-\frac{1}{2
y^3}-\frac{3}{4 y^5}+\Or(y^{-7}), \; {y\to\infty}.
\end{equation}
In view of (\ref{4-2}) it implies that $\zeta_0(s)$ does not contribute to
the heat kernel coefficients $B_k$ with $k=1/2,\,1,\,3/2$.

The analytic continuation of $\zeta_0(s)$ given in (\ref{A3}) to the
right of the strip (\ref{A3a}), i.e.\ to the region
$\mathrm{Re }\, s>3/2$, is accomplished by addition and subtraction of the
$y\to0$ asymptote
\begin{eqnarray}
 \zeta_0(s)=-\frac{\xi^2}{2}C(s)
\left\{\int\limits_{1}^{\infty}\rmd y\, y^{1-2 s}\frac{\rmd}{\rmd y}\mu_0^2(y) +\right. \nonumber\\
\left.+\int\limits_{0}^{1}\rmd y\, y^{1-2s}\left[\frac{\rmd}{\rmd y}\mu_0^2(y)
-4\left(\ln\frac{y}{2}+
\gamma+\frac{1}{2}\right)y\right]
\right\}+\zeta_{0,\rightarrow}^{\mathrm{sing}}(s)\,{,}
\label{A4a}\\
\zeta_{0,\rightarrow}^{\mathrm{sing}}(s)=2\xi^2 C(s)\left[\frac{1}{(3-2 s)^2}
+\left(\ln 2 -\gamma-\frac{1}{2}\right)\frac{1}{3-2 s}\right ] {.}
\label{A5}
\end{eqnarray}
With making use of this result we derive the contribution to the
heat kernel coefficient $B_0$ generated  by $\zeta^{\mathrm{sing}}_{0,\rightarrow}(s)$:
\begin{equation}
(4\pi)^{3/2}\mathop{\mathrm{res}}\limits_{s=3/2}\Gamma(s)\zeta^{\mathrm{sing}}_{0,\rightarrow}(s)=-4\pi\xi^2
\frac{a^2}{c^3}.
\label{A6}
\end{equation}
It should be noted that this contribution\footnote{In
\cite{BP} the calculations analogous to  (\ref{A4a}) and
(\ref{A5}) were done incorrectly. This resulted in erroneous
conclusion drawn in paper \cite{BNP}, that $B_0=0$ for a material
cylinder with $c_1=c_2$.} is due to infrared singularity of the
spectral density in the problem at hand.

The analytic continuation of the integral in (\ref{A3}) to the
left of the strip (\ref{A3a}), i.e.\ to the region  $\mathrm{Re
}\,s<-1/2$, is accomplished by addition and subtraction of the
$y\to\infty$ asymptote (\ref{A4}):
\begin{eqnarray}
\zeta_0(s)=-\frac{\xi^2}{2}\,C(s)\left[\int_{0}^{1}\rmd y \,y^{1-2 s}\frac{\rmd}{\rmd y}\mu_0^2(y)
 +\right. \nonumber\\
 \left.+\int_{1}^{\infty}\rmd y \,y^{1-2s}\left(\frac{\rmd}{\rmd y}\mu_0^2(y)+
\frac{1}{2 y^3}+\frac{3}{4 y^5}\right)
\right ]+\zeta_{0,\leftarrow}^{\mathrm{sing}}(s)\,{,}\label{A7}\\
\zeta_{0,\leftarrow}^{\mathrm{sing}}(s)=\frac{\xi^2}{2}\,C(s)
\left(\frac{1}{2}\frac{1}{1+2 s}+\frac{3}{4}\frac{1}{2 s+3}\right).
\label{A8}
\end{eqnarray}
From (\ref{A8}) we derive the respective contributions to the heat kernel coefficients
$B_{2}$, $B_{5/2}$, $B_{3}$:
\begin{equation}
\pi \xi^2\frac{c}{a^2}, \quad  0, \quad \frac{3}{4}\,\pi \xi^2\frac{c^3}{a^4}\,{.}
\label{A9}
\end{equation}

Now we turn to $\tilde{\zeta}(s)$.
To draw out the singularities in $\tilde{\zeta}(s)$ we apply the uniform asymptotic expansion of the
modified Bessel functions. In the $\xi^2$-approximation it gives
\begin{eqnarray}
\fl \ln[1-\xi^3\mu_{\nu}^2(\nu z)]= -{\xi}^2\frac{z^4 t^6}{4\nu^2}\left[1+\frac{t^2}{4 \nu^2}
(3-30 t^2+35 t^4)\right. \label{A10}\\
+\left.\frac{t^4}{4\nu^4}(9-256 t^2+1290 \, t^4 - 2037\, t^6 +1015 \,t^8)\right]
+{\Or}(\xi^4)\,{.}\nonumber
\end{eqnarray}
The integrals over $z$ converge in the range $(5-j)/2<\mathrm{Re\, }s<5/2$,
where $j$ is the power of $t$ in (\ref{A10}). After the
integration over $z$ in the divergent parts of the zeta
function $\tilde{\zeta}(s)$ we arrive at the result
\begin{eqnarray}
\fl  \zeta_{i}^{\mathrm{div}}(s)=-\frac{\xi^2  a^{2 s-1} p^{1-2 s-i}}{2\sqrt{\pi}c^{2 s}\Gamma(s)\Gamma(3/2-s)}\zeta_{\mathrm{R}}(2 s -1+i)\,
A_i(s), \quad i=1,2,...\,{,} \label{A11}\\
A_2(s)=-\frac{1}{64}\;\frac{\pi}{\cos(\pi s)} (-1+2 s)^2 (-3+2 s),\nonumber\\
A_4(s)=-\frac{1}{24576}\;\frac{\pi}{\cos(\pi s)} (-3+2 s) (27-46 s-44 s^2+56 s^3)(-1+4 s^2),\nonumber\\
A_6(s)=-\frac{1}{47185920}\;\frac{\pi}{\cos(\pi s)}(-9+4 s^2) (-1+4 s^2) \nonumber\\
\phantom{A_6(s)=}(9586 s+4640 s^5-18000 s^3+15272 s^2-2576 s^4-6345), \nonumber\\
A_{2 i+1}=0, \quad i=1,2,\ldots, \,{.}\nonumber
\end{eqnarray}
For obtaining (\ref{A11}) the integration  was carried out with
making use of the formula (\ref{3-41}).

The contributions of $\tilde{\zeta}(s)$ to the heat kernel
coefficients  $B_k$, $k=0,\,1/2,\,1,\,3/2,\,2,\,5/2,\,3$,
respectively, are
 \begin{equation}
 0, \quad 0,\quad 0,\quad \frac{3\pi^{3/2}}{8}\frac{\xi^2}{p a},
 \quad -\frac{\pi c\xi^2}{a^2}, \quad
 \frac{45\pi^{3/2}}{512}\frac{c^2 \xi^2}{p a^3}, \quad
 -\frac{3\pi}{4}\frac{\xi^2 c^3}{a^4}\,{.} \label{A11a}
 \end{equation}
Summing up the results for $\zeta_0(s)$ and  $\tilde{\zeta}(s)$,
(\ref{A6}), (\ref{A9}), (\ref{A11a}), one obtains
\[
 B_0=-4 \pi \frac{a^2}{c^2} \xi^2, \quad B_{3/2}=
\frac{3\pi^{3/2}}{8}\frac{\xi^2}{p a}, \quad B_{5/2}=
\frac{45\pi^{3/2}}{512}\frac{c^2 \xi^2}{p a^3},\]
\begin{equation}
B_{1/2}=B_1=B_2=B_3=0{.}
\end{equation}

\section{The heat kernel coefficients for a cone  with the $\delta$-potential}
In the zeta function  (\ref{3-23}) we again pick out the term with $n=0$
\begin{eqnarray}
\zeta(s)=\zeta_0(s)+\widetilde{\zeta}(s), \label{B1}\\
\zeta_0(s)=\frac{1}{2} C(s) \int_{0}^{\infty}\rmd y\, y^{1-2 s}\frac{\rmd}{\rmd y}
\ln[1+g I_{0}(y)K_{0}(y)],\nonumber\\
\widetilde{\zeta}(s)=C(s)\sum_{n=1}^{\infty}(np)^{1-2s}\int_{0}^{\infty}\rmd z\,
z^{1-2 s}\frac{\rmd}{\rmd z}\ln[1+g I_{\nu}(\nu z)K_{\nu}(\nu z)], \quad \nu= np.  \nonumber
\end{eqnarray}
In $\widetilde\zeta(s)$ we use the uniform asymptotic expansion of the modified
Bessel functions
\begin{eqnarray}
&& \ln[1+g I_{\nu}(\nu z)K_{\nu}(\nu z)]=\sum_{i=1}^\infty \frac{K_i(g,t)}{\nu^i},\quad t=
\frac{1}{\sqrt{1+z^2}}, \label{B2}\\
&& K_1=g\,\frac{t}{2}, \quad K_2=-g^2\,\frac{t^2}{8}, \quad
 K_3=g\,\left(\frac{t^3}{16}-\frac{3 t^5}{8}+\frac{5 t^7}{16}\right)+g^3\,\frac{t^3}{24},
\quad \ldots \;{.}
\nonumber
\end{eqnarray}
With (\ref{B2}), the integrals over $z$ converge in the strip
$(5-j)/2<\mathrm{Re\, }s<5/2$, where $j$ is the power of $t$, and
\begin{eqnarray}
 \widetilde\zeta^{\mathrm{div}}(s)=\sum_{i=1}^{\infty} Z_i(s), \label{B3} \\
Z_1(s)=-\frac{a^{2s-1}}{\pi c^{2s}} p^{-2s}g \zeta_{\mathrm{R}}(2s), \nonumber \\
Z_2(s)=\frac{a^{2s-1}}{2 \sqrt{\pi}c^{2s}} p^{-2s-1} g^2 \zeta_{\mathrm{R}}(2 s+1)
\frac{\Gamma(s+1/2)}{\Gamma(s)},\nonumber\\
\fl Z_3(s)=\frac{a^{2s-1}}{\pi c^{2s}\Gamma(s)} p^{-2s-2}\zeta_{\mathrm{R}}(2s+2)\left [4 g \Gamma(s+2)-
\left (g+\frac{3}{4}g^3\right )\Gamma(s+1)-\frac{4}{3} g \Gamma(s+3)\right ]{.}\nonumber
\end{eqnarray}
The contributions from $\widetilde\zeta^{\mathrm{div}}(s)$ to the heat kernel
coefficients we  calculate in the linear in $g$ approximation
by making use of (\ref{4-2}). These results are
presented in the fourth  column of  the table~\ref{table1}.

In the same approximation  $\zeta_0(0)$ is given by
\begin{equation}
\zeta_0(s)=\frac{g}{2} C(s)
\int_{0}^{\infty}\rmd y\, y^{1-2 s}\frac{\rmd}{\rmd y} \left [ I_{0}(y)K_{0}(y)\right ].
\label{B4}
\end{equation}
The  integral (\ref{B4}) converges in the region  $0<\mathrm{Re\, }s<1/2$, because
\begin{eqnarray}
\frac{\rmd}{\rmd y} [I_{0}(y) K_{0}(y)]=-\frac{1}{y}-\left(
\ln\frac{y}{2}+\gamma\right)y+\dots ,\quad y\to0; \nonumber\\
\frac{\rmd}{\rmd y} [I_{0}(y)K_{0}(y)]=-\frac{1}{2y^2}-\frac{3}{16}\frac{1}{y^4}+\dots, \quad y\to\infty.
\label{B5}
\end{eqnarray}
To perform the analytic continuation
of (\ref{B4}) to
the domain $\mathrm{Re\, }s>1/2$ one has to add and subtract from
the integrand in (\ref{B4}) several terms of its  small $y$ asymptote
\begin{eqnarray}
\zeta_0(s)&=&\frac{g}{2}
C(s)\left\{\int_{0}^{1}\rmd y\, y^{1-2 s}\left[\frac{\rmd}{\rmd y} (I_{0}(y)K_{0}(y))+\frac{1}{y}+(\ln\frac{y}{2}+\gamma)y\right]\right.\nonumber\\
&&+\left.\int\limits_{1}^{\infty}\rmd y\, y^{1-2s}\frac{\rmd}{\rmd y}I_0(y)K_0(y)\right\}+
\zeta_{0,\rightarrow}^{\mathrm{sing}}(s)\,{,}
\label{B6}
\end{eqnarray}
where the singular part
\begin{equation}
\fl \zeta^{\mathrm{sing}}_{0,\rightarrow}(s)=\frac{g a^{2
s-1}}{2\sqrt{\pi}\,c^{2s}\Gamma(s)\Gamma(3/2-s)}\left[-\frac{1}{1-2s}
-(\gamma-\ln 2 )\frac{1}{3-2s}
+\frac{1}{(3-2s)^2}\right] \label{B7}
\end{equation}
has simple poles at the points $s=3/2$ and $s=1/2$. We use
(\ref{B7}) to calculate the contribution of $\zeta_0(s)$ to the
heat kernel coefficients $B_0$, $B_{1/2}$, and $B_{1}$. These
contributions are generated by the infrared singularities of the
spectral density in the problem under consideration.

Similarly, the $y\to\infty$ asymptote (\ref{B5}) should be added
and subtracted to continue the integral (\ref{B4}) to the region $\mathrm{Re \, }s \leq 0$.
In this way we obtain
\begin{eqnarray}
\zeta_0(s)=\frac{g}{2}C(s)
\left\{\int\limits_{1}^{\infty}\rmd y\,
 y^{1-2 s}\left[\frac{\rmd}{\rmd y} (I_{0}(y)K_{0}(y))+\frac{1}{2 y^2}+\frac{3}{16}\frac{1}{y^4}\right]\right.\nonumber\\
+\left.\int_{0}^{1}\rmd y \,y^{1-2s}\frac{\rmd}{\rmd y}I_0(y)K_0(y)\right\}
+\zeta_{0,\leftarrow}^{\mathrm{sing}}(s),
\label{B8}
\end{eqnarray}
where
\begin{equation}
\zeta^{\mathrm{sing}}_{0,\leftarrow}(s)=\frac{g a^{2
s-1}}{2\sqrt{\pi}\,c^{2s}\Gamma(s)\Gamma(3/2-s)}\left(-\frac{1}{4
s}-\frac{3}{16}\,\frac{1}{2s+2}\right){.} \label{B9}
\end{equation}

According to (\ref{4-2}), the singular part (\ref{B7}) gives
contribution to the heat kernel coefficients $B_0$, $B_{1/2}$,
and $B_{1}$ while singular part (\ref{B9}) contributes to the
heat kernel coefficients $B_{3/2}$, $B_{2}$, and $B_{5/2}$.
These are given in the second column of~\tref{table1}.

Summing contributions from $\zeta_0$ and $\widetilde\zeta$, one
obtains   the heat kernel coefficients for a scalar massless
field considered on a cone with semitransparent boundary
conditions. These results  are presented in the last column of~\tref{table1}.

\begin{table}
\caption{\label{table1}The contributions to the heat kernel coefficients
$B_{k}$ coming from $\zeta_0(s)$ and $\widetilde\zeta(s)$. The
results are obtained in the linear in $g$ approximation.}
\begin{indented}
\item[]\begin{tabular}{l|rccc}\br
&$s$&$\zeta_{0}(s)\;\;$&$\widetilde\zeta(s)$& Total
\\
\mr
$B_0$&$\frac{3}{2}$& $ \pi g \frac{\displaystyle a^2}{\displaystyle c^3}$ &$0$    &$\pi g \frac{\displaystyle
a^2}{\displaystyle c^3}$
\\
$B_{1/2}$&$1$          & $ 0 $ &$0$   & $0$
\\
$B_1$&$\frac{1}{2}$& $ \frac{\displaystyle 2\pi g}{\displaystyle c} $
&$-\frac{\displaystyle 2\pi g}{\displaystyle pc}$  &$
\frac{\displaystyle 2\pi g}{\displaystyle c}\left (1-\frac{\displaystyle 1}{\displaystyle p}
\right )
$
\\
$B_{3/2}$&$0$&
$-\frac{\displaystyle 2\sqrt{\pi} g}{\displaystyle a}$ &
$\frac{\displaystyle 2\sqrt{\pi} g}{\displaystyle a}$&
$0$\\
$B_2$&$-\frac{1}{2}$& 0 &$0$&$0$
\\
$B_{5/2}$&$-1$& $ - g\frac{\displaystyle
\sqrt{\pi}c^2}{\displaystyle 2 a^3}$&  $g \frac{\displaystyle \sqrt{\pi}c^2}
{\displaystyle 2a^3} $&
$0$\\
\br
\end{tabular}
\end{indented}
\end{table}




\end{document}